\documentclass{aa}  

\usepackage{graphicx}
\usepackage{txfonts}

\usepackage{cprotect}
\usepackage{placeins}

\begin{document} 

   \title{Solar limb faculae: intensity contrast from two vantage points}

   \author{K.~Albert\inst{1}\orcid{0000-0002-3776-9548}\thanks{\hbox{Corresponding author: K. Albert.} \hbox{\email{albert@mps.mpg.de}}}
     \and
   J.~Hirzberger\inst{1} \and
   N.~A.~Krivova\inst{1}\orcid{0000-0002-1377-3067} \and
   X.~Li\inst{1}\orcid{0000-0001-8164-5633} \and
   D.~Calchetti\inst{1}\orcid{0000-0003-2755-5295} \and
   G.~Valori\inst{1}\orcid{0000-0001-7809-0067}  \and
   J.~Sinjan\inst{1}\orcid{0000-0002-5387-636X} \and
   S.~K.~Solanki\inst{1}\orcid{0000-0002-3418-8449} \and
   A.~Gandorfer\inst{1}\orcid{0000-0002-9972-9840} \and
   J.~Woch\inst{1}\orcid{0000-0001-5833-3738} \and 
   D.~Orozco~Su\'arez\inst{2}\orcid{0000-0001-8829-1938} \and
   S.~Parenti\inst{3}\orcid{0000-0003-1438-1310}
          }

  \institute{
         Max-Planck-Institut f\"ur Sonnensystemforschung, Justus-von-Liebig-Weg 3,
         37077 G\"ottingen, Germany \\ 
         \and
         Instituto de Astrofísica de Andalucía (IAA-CSIC), Apartado de Correos 3004,
         E-18080 Granada, Spain \\ 
         \and
         Institut d’Astrophysique Spatiale, Bâtiment 121, Rue Jean Dominique Cassini, Université Paris Saclay, 91405 Orsay, France
         }

   \date{Received ; accepted }

  \abstract
   {Small-scale magnetic flux concentrations contribute significantly to the brightness variations of the Sun, yet observing them~-- particularly their magnetic field~-- near the solar limb remains challenging. Solar Orbiter offers an unprecedented second vantage point for observing the Sun. When combined with observations from the perspective of Earth, this enables simultaneous dual-viewpoint measurements of these magnetic structures, thereby helping to mitigate observational limitations.}
   {Using such a dual-viewpoint geometry, we characterise the brightness contrast of faculae near the limb as a function of both their associated magnetic field strength and the observation angle.}
   {We analyse data from the Polarimetric and Helioseismic Imager on board Solar Orbiter (SO/PHI), obtained during an observation program conducted in near-quadrature configuration with Earth, in combination with data from the Helioseismic and Magnetic Imager on the Solar Dynamics Observatory (SDO/HMI). The High Resolution Telescope of SO/PHI observed a facular region located near disc centre as seen from its vantage point, while the same region was simultaneously observed near the solar limb by SDO/HMI. We identify faculae and determine their magnetic field strength from the disc-centre observations, and combine these with continuum intensity measurements at the limb to derive dual-viewpoint contrast curves. We then compare these with contrast curves derived from SDO/HMI alone.}
   {Using two viewpoints, we consistently find higher facular contrast near the limb than from a single-viewpoint. A comparison of the facular line-of-sight magnetic field 
   derived from limb observations with that derived from disc-centre observations (and re-projected to the limb) reveals large differences between the two. Co-temporal observations of limb faculae from a second, disc-centre viewpoint enable a more precise determination of their associated magnetic field.}
   {}


   \maketitle

\section{Introduction}\label{Sec:introduction}
   
In the solar photosphere, active regions appear in the form of sunspots, pores, and faculae. While large magnetic flux concentrations, such as sunspots and pores, are seen as dark features, smaller concentrations manifest as faculae, which are typically brighter than the surrounding quiet Sun and are commonly interpreted in terms of magnetic flux tubes.

The intensity of faculae relative to the local quiet Sun (hereafter referred to as intensity contrast) increases with the viewing angle from disc centre toward the solar limb \citep[e.g.][]{ortiz2002Intensity, Hirzberger2005SolarLimbFac}. This behaviour is best explained by the variation in optical depth of our observations with the angle at which we observe, which cause different portions of the flux tube walls to become visible. The walls of flux tubes appear bright due to the "hot wall effect" \citep[see,][]{spruit76}, leading to the observed intensity variation \citep[see simulations from, e.g.,][]{carlsson_observational_2004, steiner_radiative_2005}. The change of the intensity close to the extreme limb is less certain, and studies historically disagree on whether the intensity increases continuously, or if it decreases at the extreme limb \citep[see,][for an overview]{Solanki1993Smallscale}. 

More recent studies, based on full-disc observations, such as \citet{Yeo2013Intensity} and \citet{ortiz2002Intensity} find a decrease when approaching the limb, which is attributed to the surrounding granulation eventually obscuring these small structures. 
The intensity contrast of faculae at different viewing angles depends also on their associated magnetic field strength. Faculae with stronger magnetic fields generally appear brighter near the limb and darker near the disc centre than those with weaker fields. This is attributed to the correlation of the magnetic field strength to the size of the faculae. The relationship between facular intensity contrast and magnetic field strength has been examined in studies such as \citealt{kobel_continuum_2011, kahil_2019} (see also references therein), while characterising this relationship as a function of the observation angle has been addressed in works such as \citealt{ortiz2002Intensity, Yeo2013Intensity, Albert2023Faculae} -- and is the goal of the present study too. 
Beyond insight into the physics of flux tubes, these studies provide essential information for solar irradiance models spanning timescales from days to millennia \citep[see][]{krivova-2003, solanki-2013,yeo_publishers_2017,wu_solar_2018,chatzistergos-2025}.

Observing faculae, however, remains challenging, primarily because the underlying magnetic elements are frequently small ($\sim 150$\,km). In most cases, and in particular in full disc observations, they are not spatially resolved, even at the disc centre.
Consequently, the analysis of facular observations must account for the spatial averaging of signal from the flux tubes (with strong magnetic fields and high intensity contrast) and their surroundings (with considerably lower magnetic fields and intensity contrast) within each resolution element.
This challenge becomes even more severe near the solar limb. As the viewing angle increases, each resolution element covers a larger surface area due to projection effects, further reducing the effective spatial resolution. 

In addition, radiative transfer effects complicate the interpretation of observations away from disc centre. Compared to disc centre observations, at the limb, the same optical depth corresponds to higher atmospheric layers, which are generally cooler, leading to various effects. Firstly, this causes limb darkening, which decreases the signal-to-noise ratio (S/N) of limb-measurements across all Stokes parameters. 
Radiative transfer effects that influence brightness and contrast are part of the investigated phenomenon (how the intensity of faculae changes with the observation angle). However, the same effects also impact the measured Stokes profiles, and therefore the inferred magnetic quantities. This is undesirable: the magnetic field would ideally be retrieved consistently across the disc, without differences in optical depth and line-of-sight geometry, allowing to establish the variation of the intensity in relation to observation angle and the magnetic field strength coherently.
The amplitude of the Stokes V profile does not necessarily scale linearly with $\mu$, as a consequence of radiative transfer effects. Most notably, at high viewing angles, due to the narrow nature of these structures, light may pass through both magnetic and non-magnetic regions. This leads to the Stokes profiles accumulating signal from various regions \citep[see,][]{Solanki1998ReliabilityStokes, Sinjan2024Underestimate}.

In studies on solar faculae, the line-of-sight component ($B_{LOS}$) of the magnetic field is typically used as a measure of magnetic field strength due to its greater reliability compared to the full vector magnetic field. However, deriving the magnetic field strength from the $B_{LOS}$ is not straightforward. In the absence of information on the magnetic field orientation, the $B_{LOS}$ is commonly divided by the cosine of the heliocentric angle ($\mu = \cos\theta$), based on the assumption that facular magnetic fields are oriented radially with respect to the solar surface. Under this assumption, the division acts as a geometrical correction, yielding a representative magnetic field value. However,
facular magnetic fields are only radial on average; therefore, this assumed geometrical correction -- i.e., dividing the $B_{LOS}$ by $\mu$ -- introduces errors. Assuming a roughly symmetric, fanning-out configuration of field lines in facular structures, the geometrical re-projection of the field lines to the limb introduces systematic, strongly asymmetric biases in the line-of-sight component. If the same structure were observed at disc centre, the derived $B_{LOS}$ would underestimate the magnetic field strength, but no asymmetry would be introduced. See e.g., \citet{Leka2017BLOStoBr} for further discussion on using the $B_{LOS}$ to approximate the vector magnetic field, \citet{Centeno2023InversionHinodeFillingFactor} for related effects, and Calchetti et al., in prep., for effects observed in polar faculae.

Moreover, $B_{LOS}$ presents further disadvantages at the limb. Most importantly, for nearly surface-radial magnetic fields, the line-of-sight component diminishes towards the limb, leading to lower signal levels. These levels are often comparable to the noise level of the observations (which is also increased at the limb even in ideal conditions by photon noise), even for relatively strong fields, such as those associated with faculae. Moreover, limb observations probe higher atmospheric layers, where magnetic structures are more expanded, appear larger and more horizontal. The signal from these magnetic canopies can significantly contribute to the Stokes $V$ signal, altering the observed magnetic landscape \citep[see][]{GiovanelliJones1982ThreeD, Solanki1994infrared7, Pietarila2010Expansion, Ball2012TSIReconstr, Yeo2013Intensity, Albert2023Faculae}. Both of these effects hinder the reliable identification of faculae near the limb based on the $B_{LOS}$.

\begin{figure}[tbp]
   \centering
    
    \includegraphics[width=\hsize]{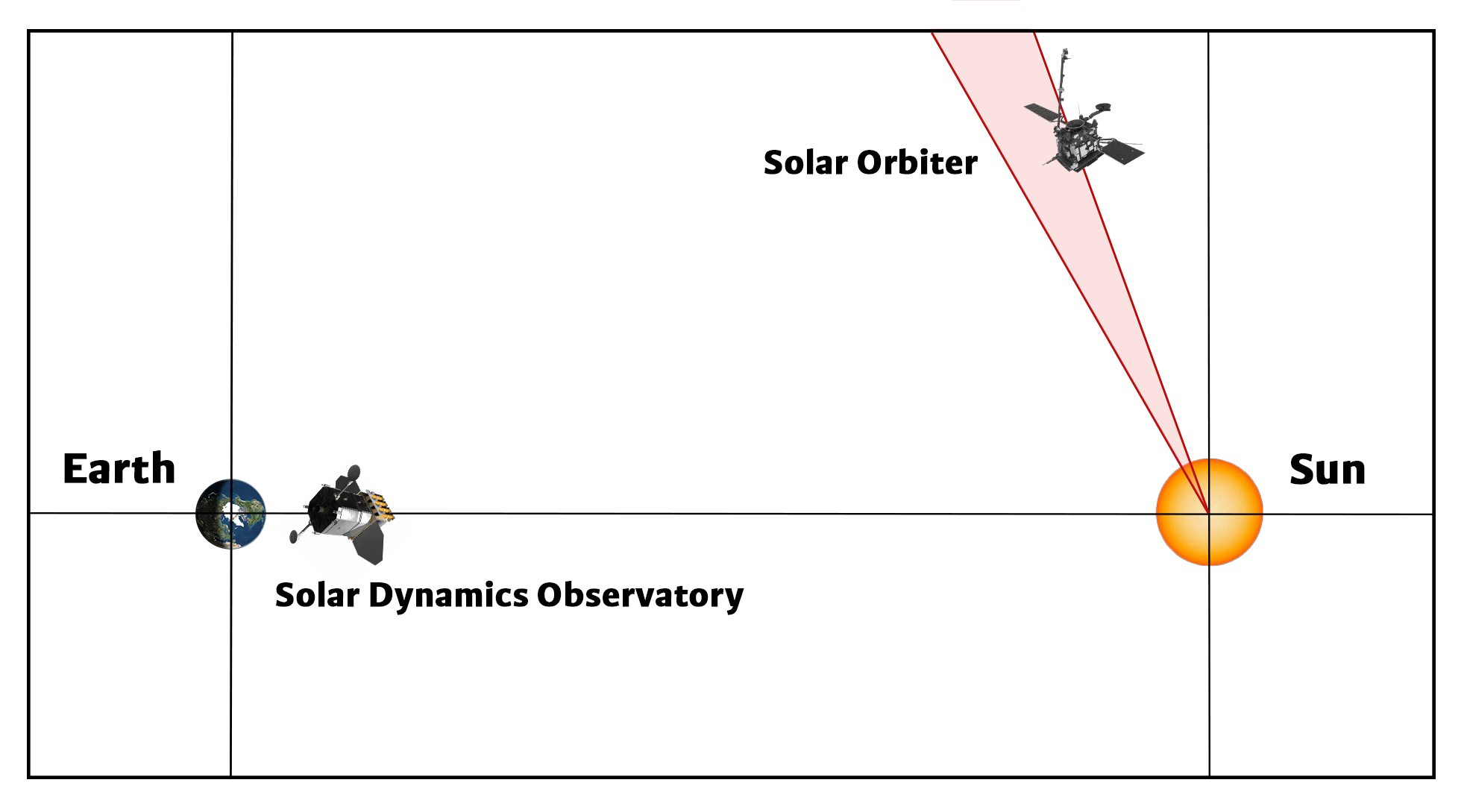}
    
     \cprotect \caption{Sketch of the observing geometry between the Solar Dynamics Observatory (SDO) and Solar Orbiter for the observations used in the study. The coordinate system is with Earth and Sun at fixed positions, as seen from above. While SDO observes the Sun from the Earth's perspective, on the horizontal black line, Solar Orbiter varies its view-angle between the two red lines, moving from right to left (i.e. approaching the Sun-Earth line). The spacecraft illustrations are a courtesy of ESA and NASA.
     }
         \label{Fig:ObservingGeometry}
   \end{figure}
   
Due to these observational limitations, the relationship between the contrast of faculae and the magnetic field strength near the limb remains uncertain in single-viewpoint studies \citep[as in, e.g.,][]{ortiz2002Intensity, Yeo2013Intensity}. A second viewpoint, where regions near the limb appear closer to disc centre, can help overcome some of these challenges. This approach was demonstrated by \citet{Albert2023Faculae}, combining data from the Full Disc Telescope of the Polarimetric and Helioseismic Imager on-board the Solar Orbiter \citep[SO/PHI-FDT,][]{muller_solar_2020, solanki_polarimetric_2020} and the Heliosismic and Magnetic Imager on-board the Solar Dynamics Observatory \citep[SDO/HMI,][]{Pesnell2012SDO, schou_design_2012}. They analysed co-observations taken at line-of-sight angles separated by $60^\circ$, and demonstrated that facular contrast can be determined much closer to the limb in the dual-viewpoint than in single-viewpoint analyses. This first study employed SO/PHI-FDT data which was acquired at a large distance from the Sun, and therefore had low resolution. 

In the current work we use high resolution observations. We analyse co-temporal observations of a faculae-rich region obtained by the High Resolution Telescope of SO/PHI \citep[HRT;][]{gandorfer18} and by SDO/HMI in a near-quadrature configuration, where the region appears close to disc centre from the perspective of SO/PHI-HRT and near the limb from the viewpoint of SDO/HMI.

\begin{figure*}[tbp]
   \centering
    \begin{minipage}{0.49\hsize}
        \centering
        \includegraphics[width=\hsize]{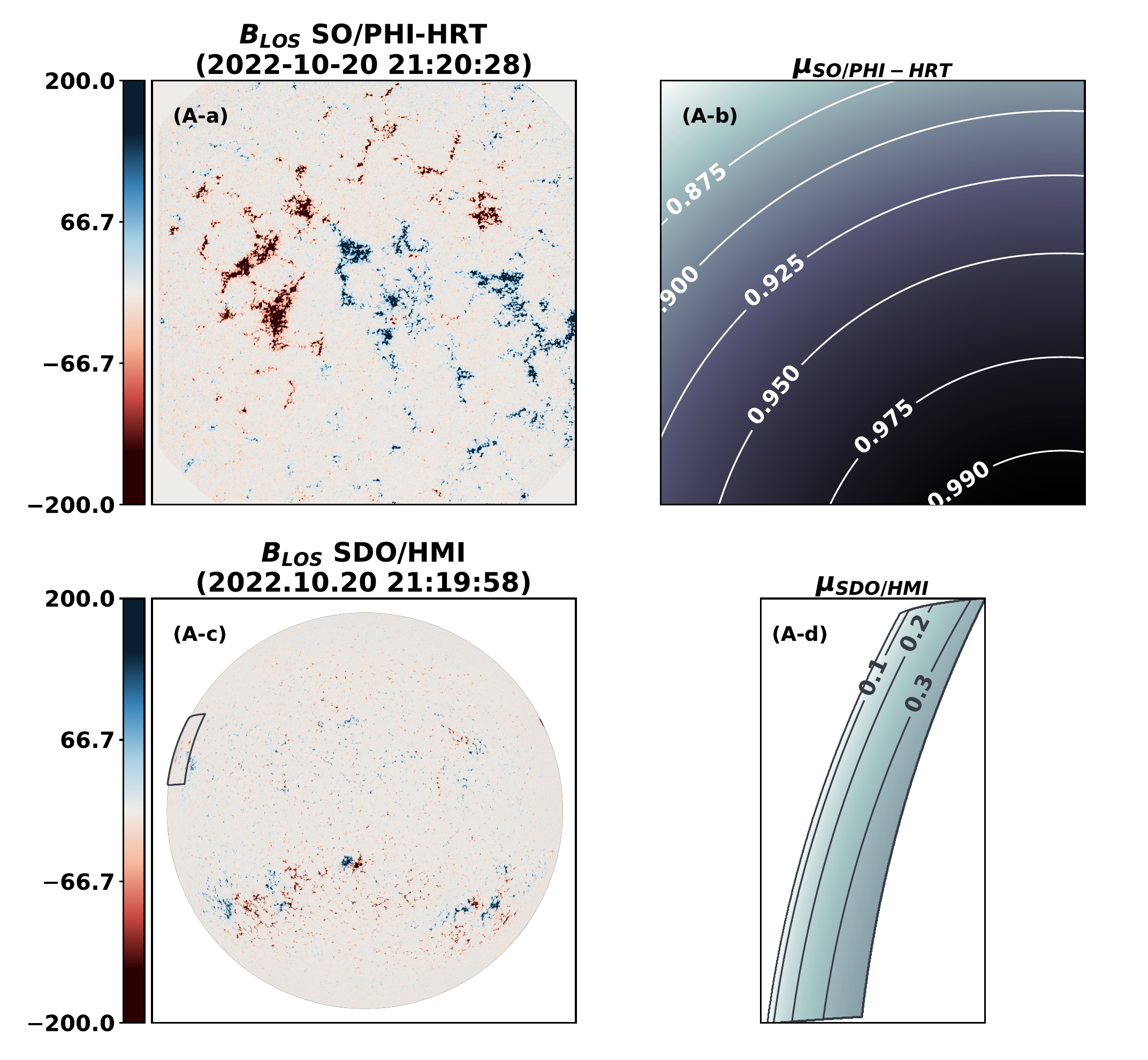}
    \end{minipage}
    \hfill
    \begin{minipage}{0.49\hsize}
        \centering
        \includegraphics[width=\hsize]{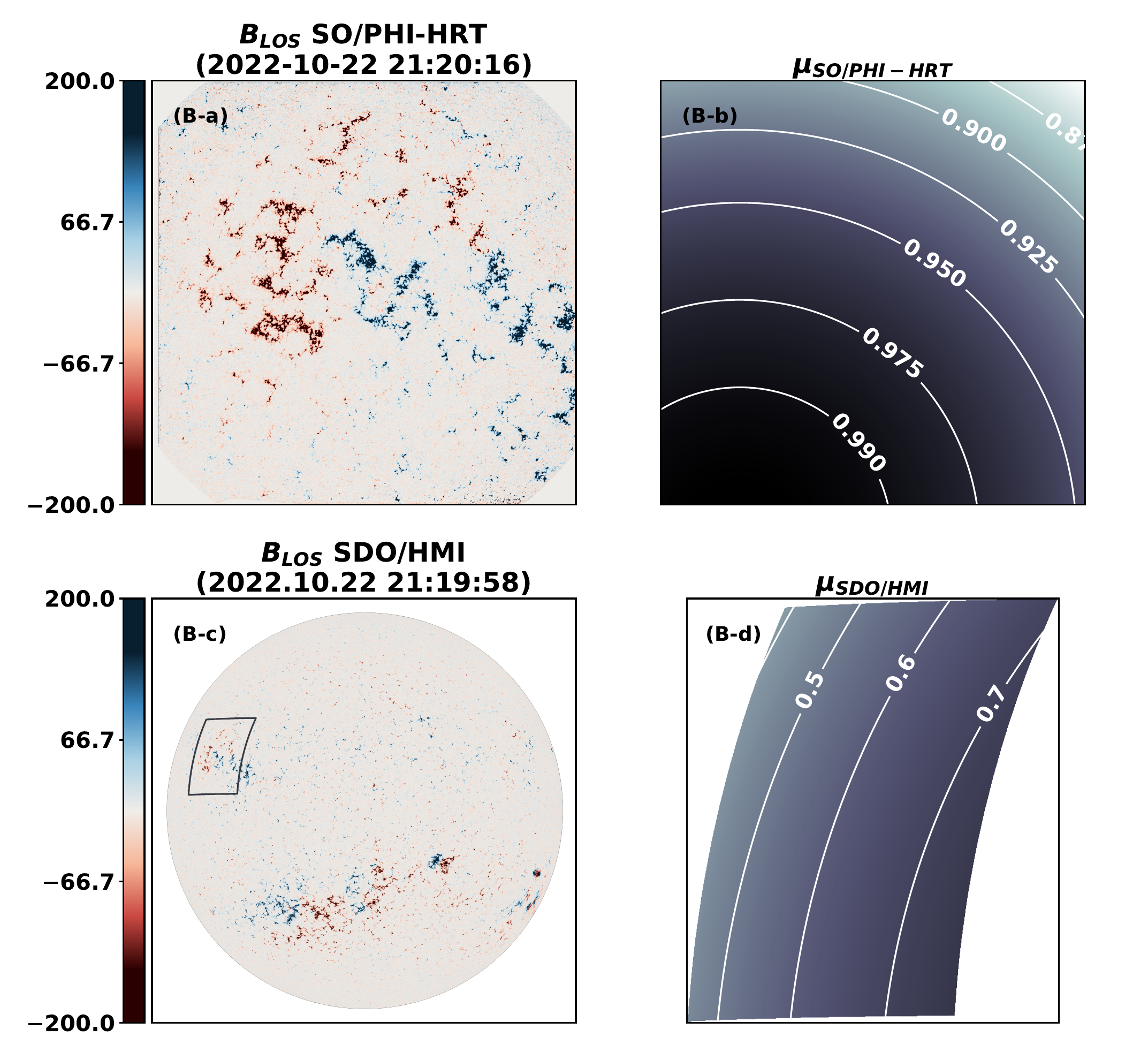}
    \end{minipage}

     \caption{The first and last data set pairs used in this study, from 20 and 22 October 2022 (A and B, respectively). Top: $B{LOS}$ and $\mu$ of a facular region near disc centre (a and b). Bottom: $B{LOS}$ and $\mu$ of the same region, as seen simultaneous by SDO/HMI (c and d). The contour shows the location of the SO/PHI field of view, and the $\mu$ in the SDO/HMI view-point is shown only for the contoured region.
     }
         \label{Fig:Snapshot_firstLast}
   \end{figure*}

\section{Methods} \label{Sec:methods} 

The data used in this study were acquired as Solar Orbiter emerged from behind the Sun from Earth's viewpoint, crossed quadrature, and approached the Sun-Earth line. During this period, the same region appeared at varying disc positions in SDO/HMI data, spanning from the extreme limb to a $\mu$ of around $0.7$. Figure \ref{Fig:ObservingGeometry} shows the observing geometry of the two spacecraft. These data provide an unprecedented opportunity to complement our understanding of facular magnetic fields at the limb through simultaneous observations of the same region from the disc-centre viewpoint.

Deriving the facular contrast as a function of magnetic field strength and observation angle (we call the resulting curves, intensity contrast curves), includes three key components: (1) the intensity contrast; (2) the magnetic field strength; (3) and the identification of faculae, which we base on the magnetic field. Our aim is to improve on some of the observational constraints at the limb, by incorporating $B_{LOS}$ obtained from a disc-centre-viewpoint into our analysis. In our study, we use (1) the contrast derived from SDO/HMI observations at the limb; (2) the $B_{LOS}$ derived at the disc centre in SO/PHI-HRT, and re-projected to the limb, resulting in spatial averaging of the SO/PHI-HRT pixels; (3) the facular map derived at the disc centre from SO/PHI-HRT, and re-projected to the limb (with the method used for the $B_{LOS}$).

The combination of intensity contrast observed at the limb with $B_{LOS}$ and facular map derived at the disc centre is only possible under the assumption that the continuum contrast only changes in magnitude when going from disc centre to the limb, but the physical area in which the contrast enhancement is observed does not become larger or smaller (other than the geometrical effect of a different view-angle). \citet{steiner_radiative_2005} found that the apparent size of faculae (i.e., the spatial extent of contrast enhancement) changes at most by $87$\,km at various view angles (measured in the plane of the sky). This is below the pixel size of SO/PHI-HRT even at the disc centre at perihelion ($\lesssim 130$\,km), and therefore the effect is negligible in our observations. 

\subsection{The observations}

We use data from a SO/PHI-HRT observational campaign in October 2022, conducted under the \verb|R_SMALL_MRES_MCAD_AR-Long-Term| Solar Orbiter Observing Plan \citep[SOOP,][]{Zouganelis2020SOOP}. At that time, from Earth's perspective, Solar Orbiter was approaching the Sun-Earth line. Observations were carried out near disc centre in the SO/PHI viewpoint (at $\mu > 0.8$), at heliocentric distances ranging from $0.35$ to $0.37$\,au. Given the $0.5"$ plate scale of the HRT, this corresponds to $127$\,km -- $134$\,km on the Sun per pixel at the disc centre, and one observation is conducted in 85\,s. Solar Orbiter tracked a region containing two largely unipolar groups of faculae, with several small pores located at the larger magnetic concentrations in both polarities.

We analyse 34 SO/PHI-HRT magnetograms from this campaign, recorded between 2022.10.20 21:15:03 and 2022.10.22 21:15:03 UTC. Specifically, we use the \verb|solo_L2_phi-hrt-blos| data product, which is processed entirely on the ground \citep[see][]{Sinjan2022HRTPipeline, Kahil2023_PSF, Bailen2023PSF}. The $B_{LOS}$ values are derived from the vector magnetic field, inferred using the \verb|cmilos| RTE inversion code \citep[see][]{Orozco2007Usefulness}. 

\begin{figure*}[tbp]
   \centering
    
    \includegraphics[width=.8\hsize]{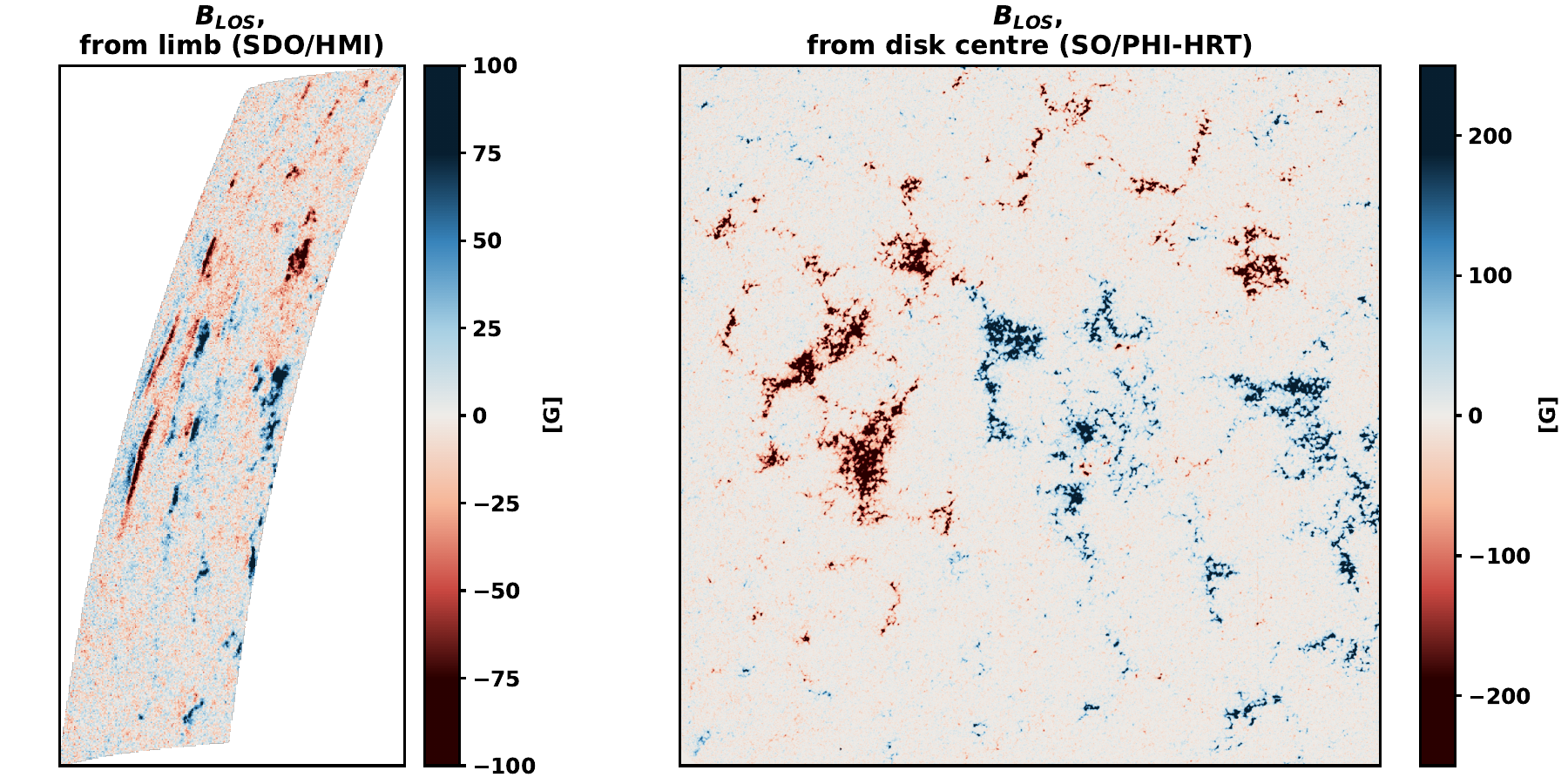}
    
     \cprotect \caption{Co-observation of $B_{LOS}$ in the same region at the limb (SDO/HMI, left), and at disc centre (SO/PHI-HRT, right), from the first data set pair in our study. The SO/PHI observation time was 2022-10-20T21:15:46, which corresponds to 2022-10-20T21:20:28 at Earth after accounting for light travel time (in UTC). The SDO/HMI observation time was 2022-10-20T21:19:58 UTC.
     }
         \label{Fig:BLOS_CompareObs}
   \end{figure*}

The SDO/HMI instrument operates on the same principle as SO/PHI. To ensure the best temporal alignment and consistency between the two instruments, we use SDO/HMI 45\,s data products. For $B_{LOS}$ we use \verb|hmi.M_45s|, computed with the so-called MDI-like algorithm \citep[named after the Michelson Doppler Imager instrument on the Solar and Heliospheric Observatory, using a similar technique, see,][]{Scherrer1995MDI}. This method computes Dopplergrams in the right and left circularly polarised light and derives the magnetic field from their difference. For continuum intensity, we use the \verb|hmi.Ic_45s| data product, which estimates the $I_c$ from the Doppler shift, line width, and line depth, assuming a Gaussian line profile \citep[see][]{couvidat_observables_2016}.

We select the SDO/HMI observations closest in time to each SO/PHI-HRT magnetogram. We do not apply any cross-calibration of the magnetic field between the instruments, as \cite{Sinjan2023_HRTCross} showed that the chosen data products agree well. Figure \ref{Fig:Snapshot_firstLast} shows the first and last dataset pairs in the data series used in this study. 
At the time of the observations, Solar Orbiter partially co-rotated with the Sun. As a result, the region remained near disc centre in the SO/PHI-HRT view, while in the SDO/HMI view it moved from close to the limb towards disc centre. Thus, the first dataset corresponds to the largest angular separation, with the region closest to the limb in HMI, while the last shows the same region closer to disc centre. 

Figure~\ref{Fig:BLOS_CompareObs} highlights the significant increase in detail in $B_{LOS}$ provided by the SO/PHI-HRT owing it to its higher spatial resolution, and its disc-centre view-point, compared to the limb view from SDO/HMI in the first data set used in our study. We exploit the high resolution of the HRT data in two ways: it allows us to reliably identify faculae, and it ensures that, even after degrading the magnetic field maps to the limb resolution for pixel-to-pixel comparison, the resulting data retain much lower noise levels.
  
\subsection{Changing geometric perspective from disc centre to limb}\label{Sec:reprojection}

We analyse pairs of datasets that capture the same region from two different viewing angles: with SO/PHI-HRT near disc centre and SDO/HMI near the limb. To enable such an analysis, we must associate each limb-pixel with a corresponding value, originating from the disc-centre observations. Besides geometrical mapping (i.e. bringing them to a common reference frame), we must also account for foreshortening -- that is, the same pixel covers an increasingly larger area on the solar surface as the observation angle increases. Additionally, the spatial resolution of the two instruments is also different. 

We change the perspective of the disc centre observation (by SO/PHI-HRT) to that of the area observed at the limb (by SDO/HMI), using the re-projection algorithm described in \citep{deforest_re-sampling_2004}, as implemented in \cite{the_sunpy_community_sunpy_2020}. This method associates each limb pixel with its footprint (i.e., contributing pixels) at the disc centre, and assigns the mean of the footprint pixels as the value of the pixel. We use the Hann window option of the algorithm for weighting the mean, which we found to provide the best balance between magnetic flux conservation and re-sampling to the lower resolution of the limb observations, specifically avoiding additional smoothing of the results. This process accounts simultaneously for both the foreshortening effect caused by observing the same area from different viewing angles and the difference in spatial resolution between the two instruments, and we do not apply prior spatial degradation to the SO/PHI data.  We use the term re-projection, following \citep{deforest_re-sampling_2004}, but we note, that there is no vector re-projection performed, rather just a re-mapping or re-sampling of the pixels.

For the reference frame used in the geometrical mapping, we rely on the georeferencing of the data sets \citep[expressed in the World Coordinate System framework, WCS, see][]{Thompson2006WCS}. For full-disc observations, such as those from SDO/HMI, we can take the disc centre as a reference, which can be determined with high accuracy by using limb-fitting algorithms. In contrast, the SO/PHI-HRT data used in the study lacks clear reference features like the solar limb within its limited field of view. Consequently, its WCS coordinates rely on the spacecraft's attitude determination, which is less precise. To improve the accuracy of the SO/PHI-HRT WCS coordinates, we first re-project the SDO/HMI data to the SO/PHI reference frame and perform a cross-correlation between the two. The resulting shift is then used to update the WCS entries to the SO/PHI-HRT data.

After updating the WCS coordinates of the SO/PHI-HRT data, we apply image distortion corrections and re-project the data into the SDO/HMI reference frame. To identify residual misalignment, we compute the cross-correlation between the re-projected SO/PHI $B_{LOS}$ and the $I_c$ provided by SDO/HMI near the limb. We refine the alignment to sub-pixel accuracy by fitting a quadratic surface to the $3 \times 3$ pixel region surrounding the correlation peak and calculating its maximum. This ad hoc alignment assumes that the disc-centre and limb views correspond to the same solar region without any significant parallax effect. 

\subsection{Viewing angle compensation of $B_{LOS}$}

Due to the evacuated nature of magnetic flux tubes, faculae are buoyant in the surrounding granulation and, on average, they exhibit surface-vertical orientation
\citep[see, e.g.,][]{KnolkerSchussler1988FluxTubeModel, Schussler1992SmallScaleMagneticFieldsBook}. 
Therefore, the line-of-sight component of the magnetic field, $B_{LOS}$, decreases toward the limb. This is commonly compensated by dividing $B_{LOS}$ by $\mu$, which is a simple geometrical correction under this surface-radial assumption \citep[see also,][]{Murray1992InclinationOfFields}.

However, facular structures are not strictly radial, but expand with height \citep[][]{Solanki1999Expansion}. In observations of faculae near the limb, a reversal in magnetic polarity is observed for the limb-ward sides of facular regions \citep[see][]{Pietarila2010Expansion}. This is due to the fact that the inclined fields at observation angles larger than their inclination, tilt away from the observer, and show a reversal of their polarity in the $B_{LOS}$. These observations indicate that the structures in facular regions are sufficiently inclined, to introduce strong asymmetric biases in the derived $B_{LOS}$ at large heliocentric angles. Meanwhile, we can expect that approximating the magnetic field strength with $B_{LOS}/\mu$ at or close to the disc centre underestimates the magnetic field due to the inclined field lines, but does not introduce strong asymmetries.
By incorporating simultaneous disc-centre observations from a second viewpoint, we can determine the magnetic field strength of facular structures near the limb without the strong biases affecting limb measurements, and we can ensure consistency across all viewing angles.

\subsection{Continuum intensity contrast}\label{Sec:contrast}
The continuum intensity contrast quantifies the brightness of facular features relative to their local surroundings. In this study, we use the contrast derived exclusively from the SDO/HMI viewpoint.

To begin, we calculate the normalised intensity, $I_{c, norm.}(x,y,t)$, by dividing the continuum intensity images by the local $\mu$-dependent mean quiet Sun intensity, where $x$ and $y$ are the spatial dimensions, and $t$ denotes time. This relationship is similar to the commonly used centre-to-limb variation (CLV) of solar intensity, with the distinction that it accounts not only for limb darkening but also for stray light and other instrumental artefacts. We determine this relationship in the individual data sets by computing the mean intensity of the quiet Sun regions (defined as regions where $|B_{LOS}|<2*\sigma$, $\sigma$ being the noise level of SDO/HMI) as a function of $\mu$. We sort the $I_c$ data by $\mu$, and group them into bins of $5000$ pixels. The mean intensity in each bin is then fitted with a $20$th-order polynomial, which provides an accurate representation of the intensity variation for $\mu > 0.035$ (see Appendix~\ref{App:A} for further details).

The continuum intensity contrast, $C_{I_\text{c}}(x,y,t)$, is then calculated as:
\begin{equation}
    C_{I_\text{c}}(x,y,t) = I_{c,\text{norm}}(x,y,t) - 1.
    \label{Eq:Int_contr}
\end{equation}

\begin{figure*}[ptb]
   \centering
    
    \includegraphics[width=\hsize]{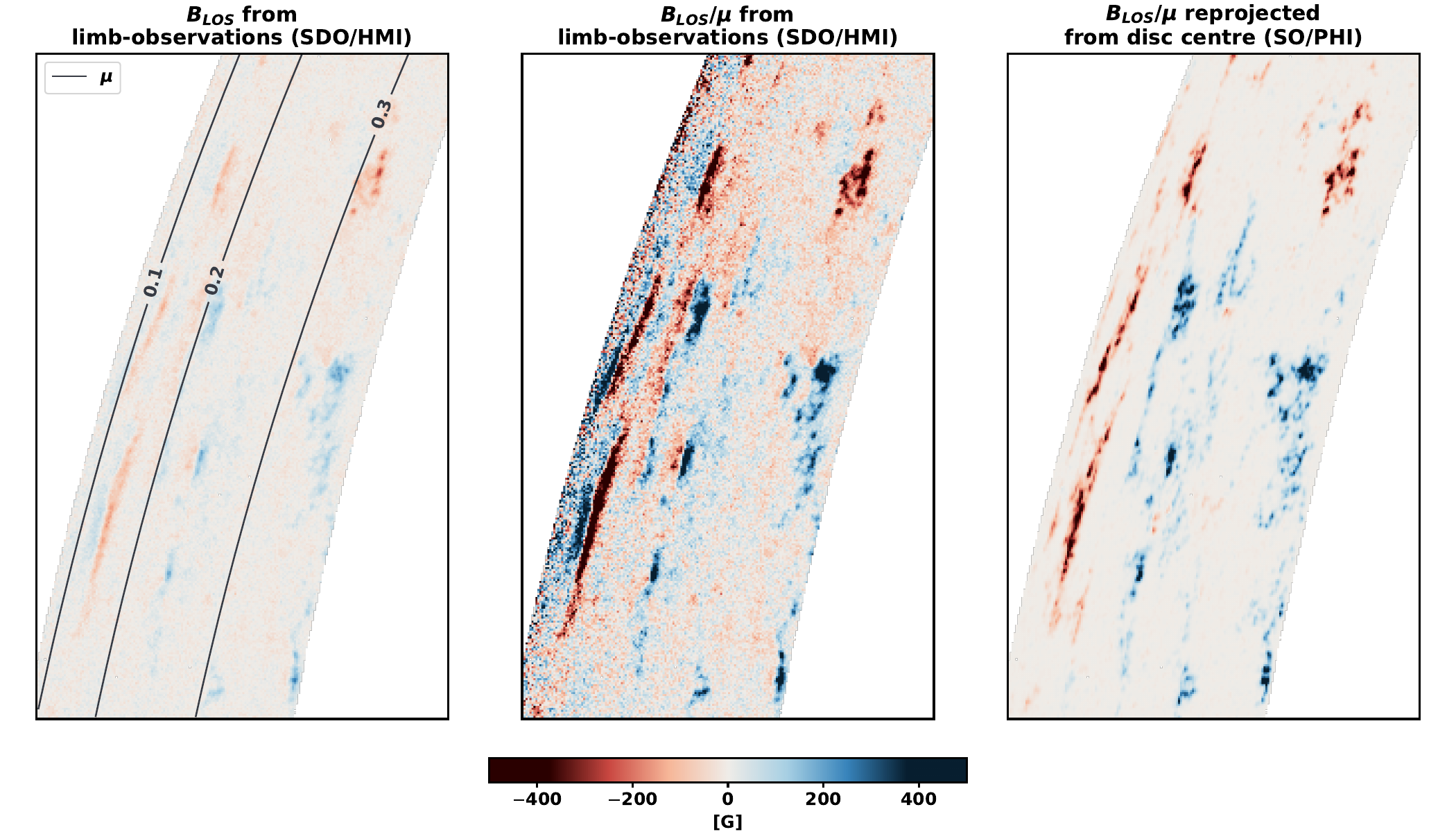}
    
     \cprotect \caption{Sub-region of the first dataset pair in the time series used in the study, showing: $B_{LOS}$ derived from the limb observation by SDO/HMI (first panel); the same data divided by $\mu$ ($B_{LOS}/\mu$, second panel), and $B_{LOS}$ derived from disc centre by SO/PHI-HRT, divided by the $\mu$ value of the disc-centre view and re-projected to the limb to match the reference frame of SDO/HMI (third panel).
     }
         \label{Fig:BLOS_compared}
\end{figure*} 

\subsection{Identification of faculae}\label{Sec:facularMap}
We identify faculae, based on their associated $B_{LOS}$ values, following the approach of \cite{Albert2023Faculae}, applied in the respective observed coordinate systems. The result of this identification is referred to hereafter as the "facular map". A pixel is classified as facular if it is part of a contiguous structure, exceeds three times the noise level in $B_{LOS}$, and is not dark (i.e., not a sunspot or pore, assessed based on an intensity threshold).
We confirmed the robustness of the three-times-noise-level threshold by increasing it to four, which yielded no significant change in the results.

For SDO/HMI, the noise level follows a centre-to-limb variation, which we determine using the method described in \cite{Albert2023Faculae}. However, in contrast to \cite{Albert2023Faculae}, we do not account for field-of-view dependent residual variations in the noise additional to the centre-to-limb changes. We note, that the noise levels at the East and West limbs may deviate by up to $1$\,G from the mean noise level due to the radial velocity of SDO (private communications A. Kumar Yadav), however we do not account for these variations. The derived noise level ranges from $10.2$\,G near the limb to $8.6$\,G at disc centre. These values are largely consistent with those reported by \cite{Liu2012CompBLOS} at disc centre, although they are somewhat lower at the limb.

\begin{figure*}[tbp]
   \centering
    
    \includegraphics[width=\hsize]{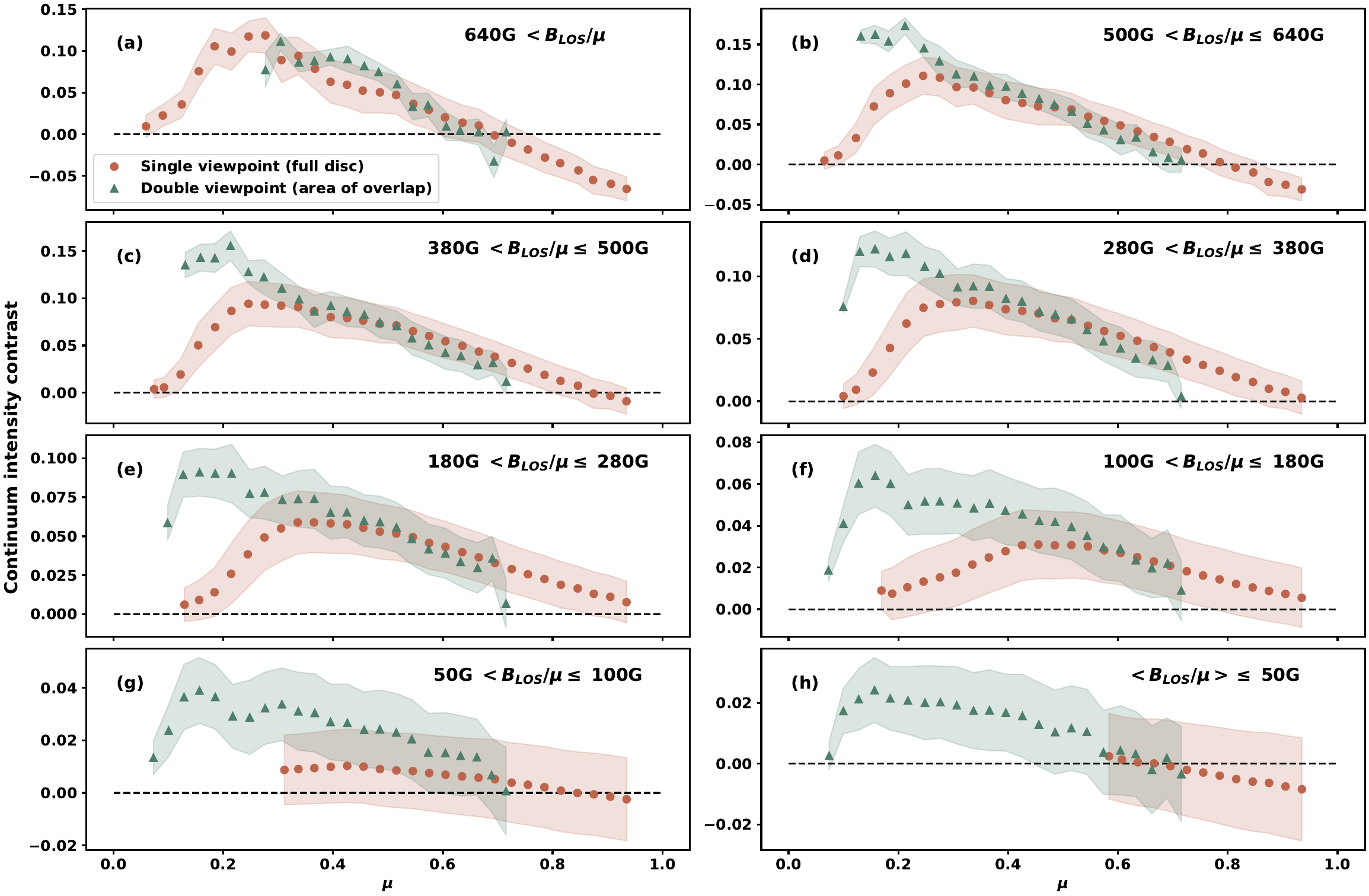}
    
     \caption{Relationship between facular intensity contrast and $\mu$. Panels (a)-(h) show pixels grouped by their associated $B_{LOS}/\mu$ values. Points represent the mean intensity contrast in $\mu$ bins of width 0.03; shaded areas indicate the standard deviation within each bin. Orange dots correspond to a single-viewpoint analysis using SDO/HMI data from the full disc. Green triangles represent results from the dual-viewpoint approach, combining SO/PHI-HRT observations at disc centre with SDO/HMI observations near the limb in the region of overlap. Note that the y-axis limits differ between panels.
     }
         \label{Fig:CurvesVsMu}
  \end{figure*}

In addition to the $B_{LOS}$ threshold, we apply an intensity-based criterion to exclude sunspots and pores in the full disc SDO/HMI data, following \citet{Albert2023Faculae}. Specifically, any pixel with a continuum intensity below a certain threshold is excluded. This threshold is defined as the mean of the minimum values of $I_{c, norm.}(x,y,t)-3\sigma_{I_c,QS}(x,y,t)$ across all observations, where $\sigma_{I_c,QS}$ is the standard deviation of the quiet Sun continuum intensity (i.e., of those pixels that did not classify as faculae). Note, that this facular map is used only for the reference results from a single-viewpoint, for which we use SDO/HMI data from the full disc (see Fig. \ref{Fig:CurvesVsMu}).

For SO/PHI, we assume a uniform $B_{LOS}$ noise level of $10$\,G across the entire field of view. Since the observed region at disc centre contains no sunspots or large pores -- except for small pores, which appear bright near the limb, and hence also contribute to the facular brightness there -- we do not apply additional intensity-based selection criteria.

The facular map determined from SO/PHI is re-projected to the SDO/HMI coordinate system. Since we map data from disc centre to the limb, the projected area becomes compressed, meaning that multiple disc-centre pixels -- with potentially not all being classified as the same type of structure -- contribute to a single pixel at the limb. As a result, the re-projected facular map contains pixels with varying degrees of facular contribution. To ensure that only pixels with sufficient facular contribution are included in our analysis, we apply an additional filtering criterion: any pixel for which the re-projected $B_{LOS}/\mu$ from SO/PHI-HRT falls below the SDO/HMI $B_{LOS}$ noise level at the limb is excluded \citep[see also][]{Albert2023Faculae}. The excluded pixels typically lie at the peripheries of facular regions or correspond to small, scattered flux concentrations, consistent with the effects of reduced spatial resolution (see Fig. \ref{Fig:Faculae_reproj} in Appendix \ref{App:B}).

\section{Results}
  \begin{figure}[tbp]
   \centering
    
    \includegraphics[width=\hsize]{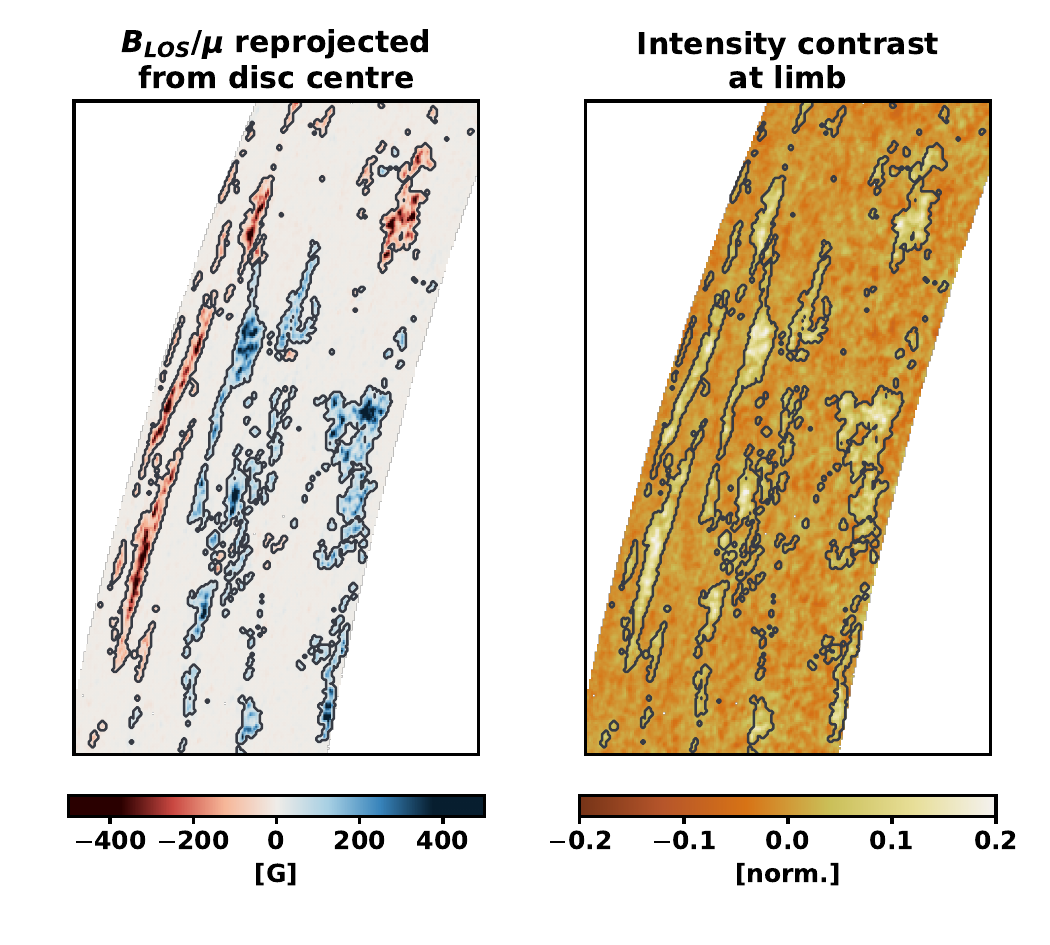}
    
     \caption{The data combined in our analysis from different viewing angles. Left: $B_{LOS}$ derived at disc centre from SO/PHI-HRT observations, normalised by the local $\mu$ at which it was observed, and reprojected to the limb. Right: the intensity contrast is computed from SDO/HMI $I_c$ data at the limb. Contours: we identify facular pixels at disc centre in SO/PHI-HRT observations, then re-project this map to the limb. The same sub-region is shown as in Fig. \ref{Fig:BLOS_compared}.}
         \label{Fig:inputToAnalysis}
   \end{figure}

\subsection{Comparison of the two viewpoints} \label{Sec:resultsViewPoints}
We begin by comparing the $B_{LOS}$ and $B_{LOS}/\mu$ values derived at the limb from SDO/HMI with the $B_{LOS}/\mu$ of the same region, derived from disc centre co-observations by SO/PHI and re-projected to the limb. In Fig. \ref{Fig:BLOS_compared}, we show a sub-region of the first dataset pair used in our analysis, where the observed area lies closest to the limb in SDO/HMI ($\mu<0.4$).

A key difference between the two viewpoints is that the same region shows patches of largely unipolar $B_{LOS}$ at the disc centre, while it exhibits a polarity reversal at limb -- similar to what is observed for sunspots. 
This behaviour is consistent with previous findings by \citet{Pietarila2010Expansion}, and here we confirm these for the first time from an independent vantage point. The underlying cause is the expansion of the magnetic structures with height, which leads to the inclination of the magnetic field lines relative to the surface normal \citep[as also modelled in][]{Pietarila2010Expansion}. On the disc-centre facing side of a fanned-out structure, the field lines point toward the observer, maintaining the same polarity as observed at disc centre. On the limb-facing side, however, the field lines tilt away from the observer, resulting in the reversal of their polarity in the observed $B_{LOS}$ \citep[see also,][]{GiovanelliJones1982ThreeD, Solanki1994infrared7}. This also leads to the appearance of a polarity reversal line, with close-to-zero $B_{LOS}$ values in the magnetograms at the limb.

Including a simultaneous disc-centre viewpoint of the limb faculae offers two key advantages in the analysis. First, deriving the magnetic field of limb faculae from the disc-centre viewpoint reduces the biases present in limb observations (such as geometrical projection and foreshortening effects, detailed in Sec. \ref{Sec:introduction}). This ensures a more consistent determination of the facular magnetic field across the solar disc. Second, because in this study faculae are identified based on their associated magnetic field (see Sec. \ref{Sec:facularMap}), a more accurate field determination leads to a more reliable facular map at the limb.

\subsection{Intensity contrast curves}\label{Sec:resultsCurves}
   
\begin{figure}[tbp]
   \centering
    
    \includegraphics[width=\hsize]{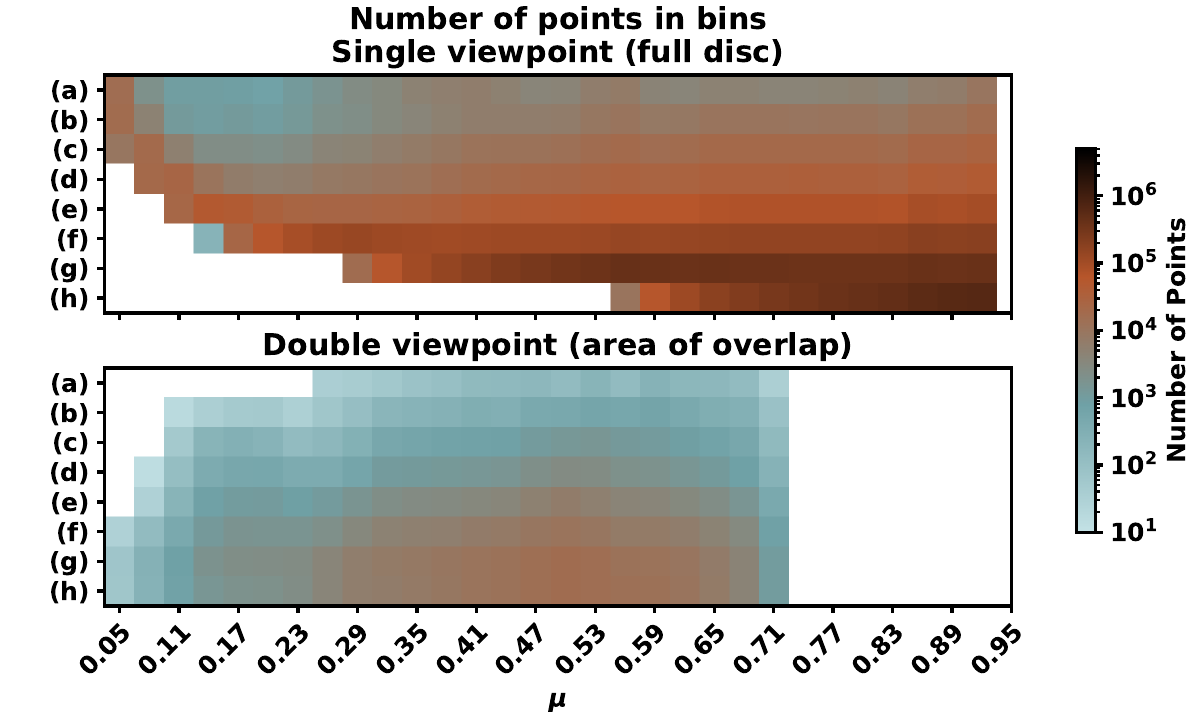}
    
     \caption{Number of pixels averaged in each $\mu$ bin for the results in Fig. \ref{Fig:CurvesVsMu}. The y-axis is labelled "a" to "h" in accordance with the panel identifiers in Fig. \ref{Fig:CurvesVsMu}. Top panel: single-viewpoint case (orange dots in Fig.~\ref{Fig:CurvesVsMu}). Bottom panel: double-viewpoint case (green triangles in Fig.~\ref{Fig:CurvesVsMu}). A logarithmic colour scale is used. Note, that the single-viewpoint analysis has more pixels overall, as we consider the entire solar disc, as opposed to the small region of overlapping field-of-view in the double-viewpoint.
     }
         \label{Fig:PixelStatistics}
  \end{figure}
  
Figure~\ref{Fig:inputToAnalysis} illustrates the combination of data used in the double-viewpoint analysis, showing a sub-region of the first data pair included in the study. The data is processed as described in Sec.~\ref{Sec:methods}.

To characterise the $C_{I_c}$ of faculae as a function of both the associated $B_{\mathrm{LOS}}/\mu$ and $\mu$, we derive intensity contrast curves ($C_{I_c}$ versus $\mu$) for different $B_{\mathrm{LOS}}/\mu$ categories (Fig.~\ref{Fig:CurvesVsMu}). Faculae are grouped into eight categories based on their $B_{\mathrm{LOS}}/\mu$ values. Within each category, data are binned in intervals of $\Delta \mu = 0.03$. Only points with $\mu \geq 0.05$ are considered, and any bin with fewer than 10 points is disregarded. 

Figure~\ref{Fig:CurvesVsMu} presents the centre-to-limb variation of the facular continuum contrast. This is shown separately for eight bins in $B_{\rm LOS}/\mu$ in the different panels. In each panel two curves are plotted: a reference derived from a single viewpoint (from SDO/HMI, shown in orange), and the results of our double-viewpoint approach (combining SDO/HMI and SO/PHI-HRT in their overlapping region, shown in green). The markers represent the mean $C_{I_c}$ within each bin, while the shaded regions indicate the standard deviation.

\begin{figure*}[ptb]
   \centering
    
    \includegraphics[width=\hsize]{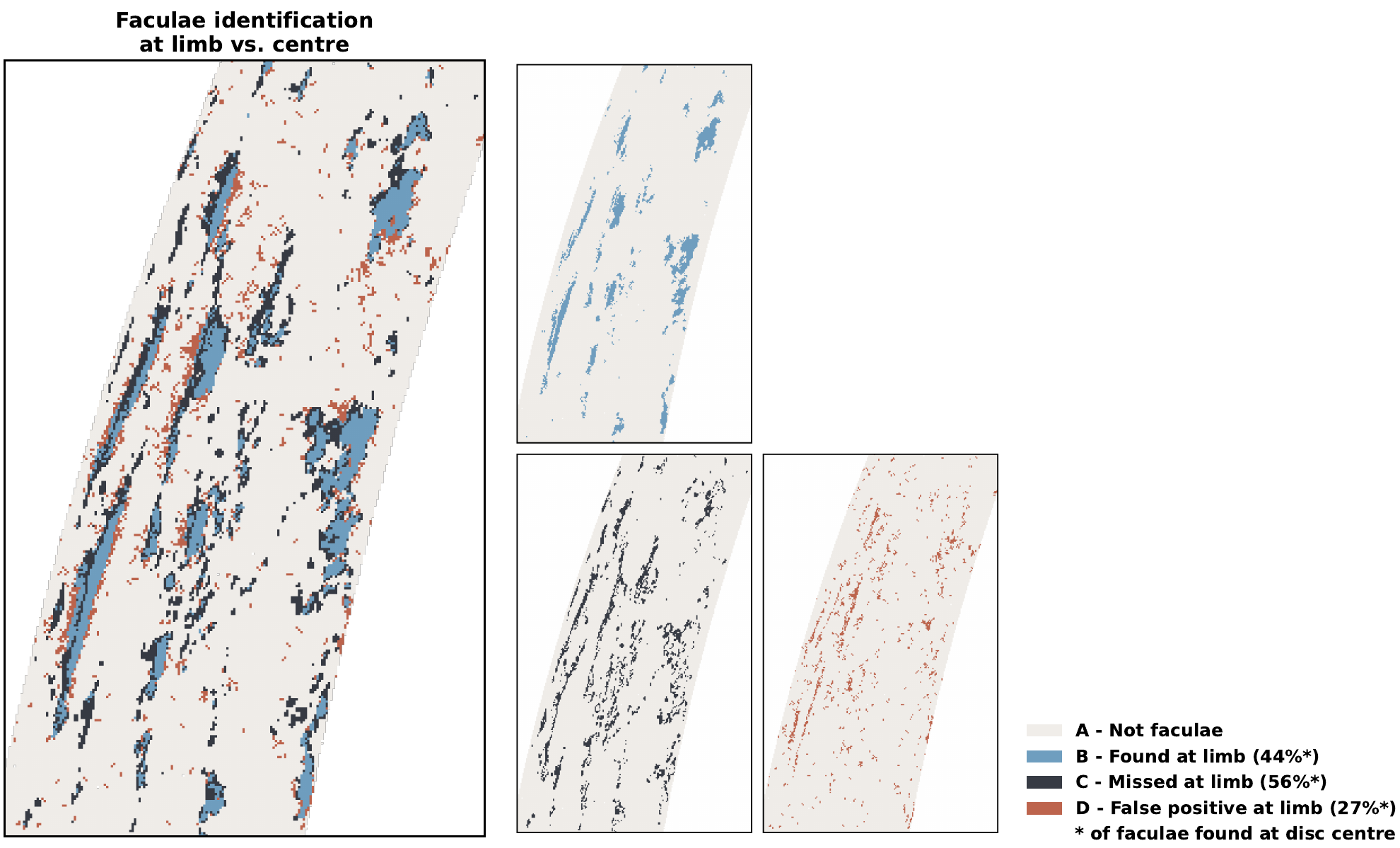}
    
     \caption{Differences in faculae identification between direct limb observations (as in state-of-the-art studies) and identifications performed in disc-centre data, with the results reprojected to the limb (this study). Colour B shows facular pixels identified both at disc centre and at limb, while colour C shows the faculae missed at the limb compared to the disc centre view. Colour D shows pixels identified as faculae in the limb data, which do not appear as faculae at the disc centre. The same sub-region and dataset pair as in Fig. \ref{Fig:BLOS_compared} are shown.
     }
         \label{Fig:comparison_FacMap}
\end{figure*}

The orange curves are derived from the full solar disc observed in SDO/HMI, and therefore are comparable to the results of \citet{Yeo2013Intensity}. There are two major differences in how the SDO/HMI data is treated in \citet{Yeo2013Intensity} and our study. \citet{Yeo2013Intensity} (1) averaged seven consecutive \verb|hmi.M_45s| data products, and (2) used a total of $15$ data sets, spanning over a period longer than a year. In contrast, in this study we (1) use single \verb|hmi.M_45s| magnetograms, and (2) our statistics are based on $34$ data sets, recorded in a time-span of two days. The largest difference in the results is that the noise level was lower in the averaged magnetograms, leading to the curves extending closer to the limb in \citealt{Yeo2013Intensity}. Apart from this aspect, the results agree well.

The green curves, derived using the double-viewpoint approach, show that for faculae associated with lower magnetic field strengths ($B_{\mathrm{LOS}}/\mu < 180$\,G), the method enables intensity contrast observations significantly closer to the limb than the single-viewpoint approach \citep[as also demonstrated by][]{Albert2023Faculae}. Additionally, we find few faculae linked to strong magnetic fields ($B_{\mathrm{LOS}}/\mu > 640$\,G) near the limb in the double-viewpoint results, showing up as the green curves not extending all the way to the limb in panels (a) to (c). This is a consequence of the magnetic field now being derived at the disc centre, and therefore showing lower values (see also Sec. \ref{Sec:resultsViewPoints}, and Appendix \ref{App:A0}). These contrast curves also exhibit trends distinct from those of the single-viewpoint curves. In the single-viewpoint case, the contrast generally increases smoothly from disc centre toward the limb before gradually decreasing \citep[as also observed by][]{ortiz2002Intensity, Yeo2013Intensity}. In comparison, the double-viewpoint results show a close-to-linear increase in contrast toward the limb, followed by a sharp drop. 

\begin{figure}[tbp]
   \centering
    
    \includegraphics[width=.9\hsize]{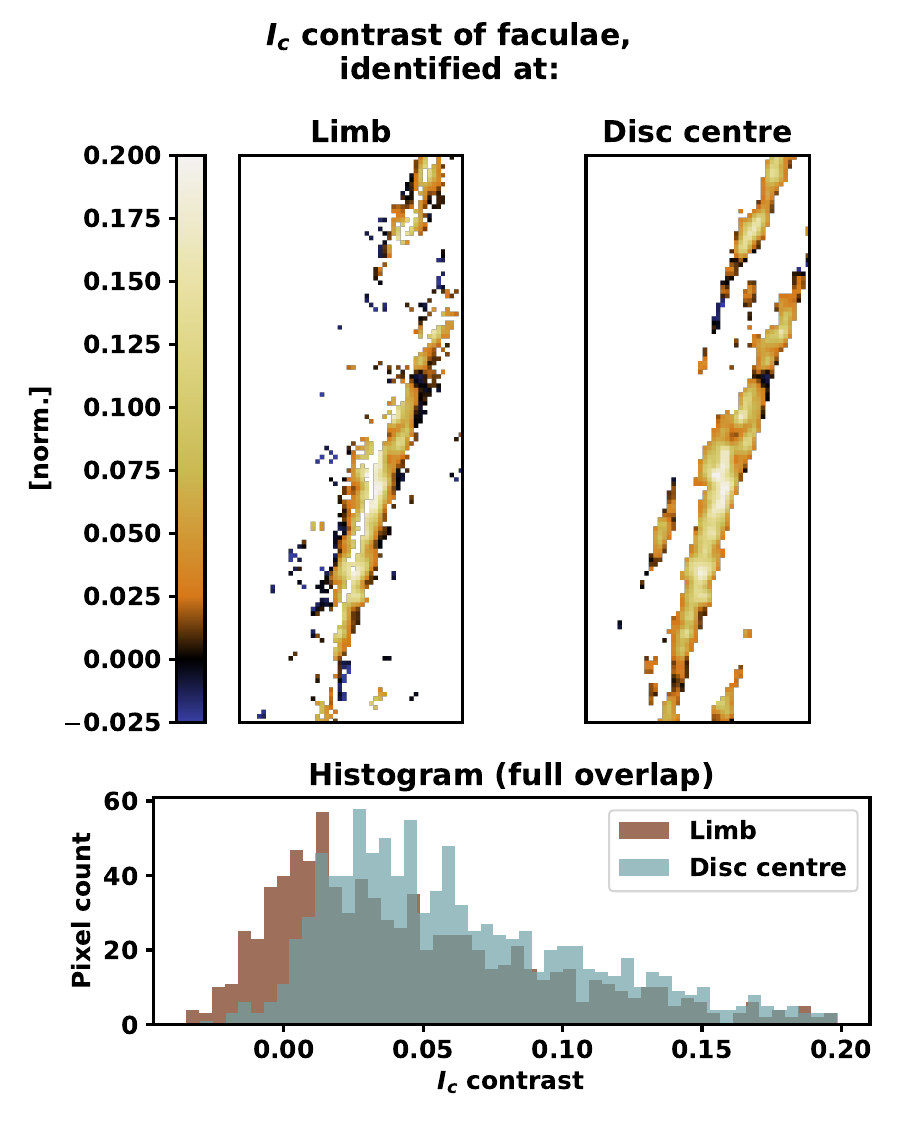}    

     \caption{Intensity contrast of a facular region at the limb. Left: facular pixels are identified at the limb. Right: facular pixels are identified at the disc centre. Bottom: histogram of the pixels in the top left and right panels. We show a sub-region of the first dataset in the study, corresponding to $\mu<0.2$.}
         \label{Fig:Ic_identification}
\end{figure}

Figure~\ref{Fig:PixelStatistics} shows the pixel counts in each bin for both single- and double-viewpoint methods. In the single-viewpoint case, stronger magnetic fields are predominantly detected near the limb, producing an artificial distribution in which high-field faculae appear more frequently at low $\mu$ values than closer to disc centre. 
This contradicts the expectation that $B_{LOS}$ measured at lower spatial resolution (larger pixels near the limb) should be reduced by averaging with neighbouring signal, rather than enhanced relative to the higher-resolution disc-centre view. In contrast, the double-viewpoint method produces a more balanced distribution, with most pixels concentrated at intermediate $\mu$ values and gradually decreasing toward the extremes. This pattern aligns with the motion of Solar Orbiter during the observations, moving from quadrature towards the Sun-Earth line, corresponding to $0\leq\mu\leq0.75$ in SDO/HMI. Consequently, the overlapping area of most data sets -- and thus the region with increased statistics -- is centred around intermediate $\mu$ values of the observed range. With this new distribution, however, we have low statistics in the higher magnetic field categories, as well as in regions close to the limb.

The observation angle at which facular contrast reaches its maximum ($\mu_{max}$), has long been debated \citep[see][]{Solanki1993Smallscale}, and observations disagree due to differences in, for example, observational resolution and facular identification methods \citep[see][]{Centrone2003CLVFac_ident, AuffretMuller1991CLV_NBP}.
In our results, the maximum contrast lies systematically closer to the limb, especially in the lower magnetic field cases (panels d-h), than observed from a single view-point. However, especially in panels (a)-(c), the statistics are too low to form a clear peak. While in panels (d)-(h) we can identify a clear peak, this lies so close to the limb, that the decrease after the peak is defined only by a few points, which all have relatively low statistics. Therefore, due to the limited statistics, we do not fit a curve to our results to determine the $\mu_{max}$ \citep[as in, e.g.,][]{Albert2023Faculae, Yeo2013Intensity}.

\section{Discussion}

The observed differences between the single- and double-viewpoint contrast curves in Fig.~\ref{Fig:CurvesVsMu} arise from two main factors: (1) the pixels in the analysis are selected based on the new facular map derived at disc centre, and (2) the magnetic field values associated with the faculae are likewise taken from disc-centre observations. Since faculae are identified via their associated $B_{LOS}$ values, both effects ultimately reflect the differences between the magnetic landscapes seen at the limb and at disc centre (Fig.~\ref{Fig:BLOS_compared}).

Figure~\ref{Fig:comparison_FacMap} highlights the discrepancies between the facular maps derived from limb data (based on SDO/HMI), and disc centre data (based on SO/PHI) re-projected to the limb. Using the disc-centre identification as a reference, only $44\%$ of faculae are correctly identified in the data gathered near the limb, while $56\%$ are missed -- mainly along the edges of the facular regions, although some of the smaller facular structures are entirely undetected. There are also a number of pixels identified as faculae at the limb, which do not classify as faculae from the disc centre. These are mainly along the edges of the facular structures, with some smaller concentrations scattered around the region. The latter could be the signature of horizontal magnetic fields at the extreme limb \citep[see, e.g.,][]{Lites2008HorizontalFields, Danilovic2010TransversePhotosphere}. Figure~\ref{Fig:Ic_identification} shows the intensity contrast of the pixels identified as faculae at the limb, and at the disc centre. The disc-centre identification outlines the contrast enhancement better, while the limb-identification includes many dark pixels and misses bright pixels in the centre of the structure, also shown in the histogram.

 \begin{figure}[tbh]
   \centering
    
    \includegraphics[width=.9\hsize]{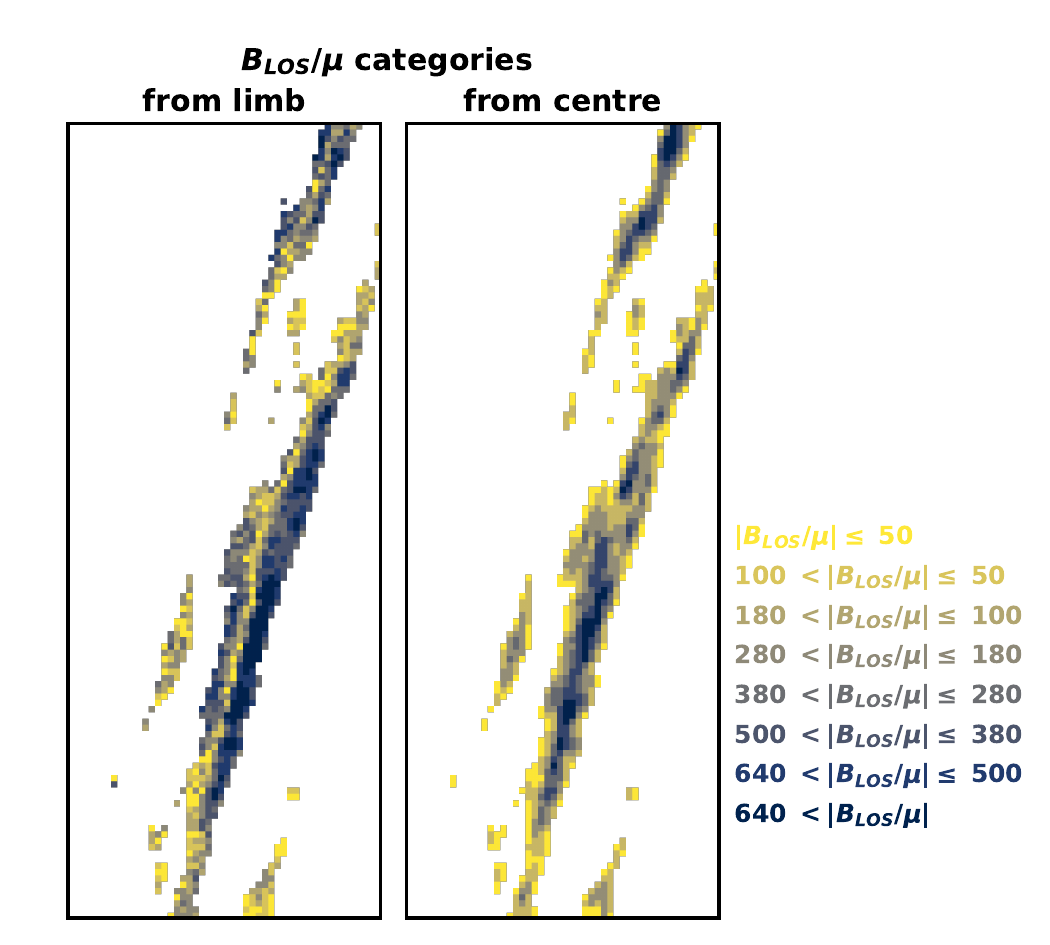}
    
     \caption{Sub-region of the first dataset pair in the study, illustrating the change in $B_{LOS}$ category of facular pixels between vantage points. Left: limb-view. Right: disc centre-view. The categories (a to h in Fig.~\ref{Fig:SameRegion}) are assigned colours ranging from blue to yellow, which are used to colour the pixels in the images. Same region as in Figure \ref{Fig:Ic_identification}.
     }
         \label{Fig:Recategorisation}
  \end{figure}

Using the magnetic field derived at disc centre instead of the limb causes pixels to be reclassified among the different magnetic field categories. Figure~\ref{Fig:Recategorisation} illustrates an example of how the $B_{LOS}/\mu$ categories differ between centre and limb views. Pixels are coloured according to their assigned category. 
In the limb view, pixels falling into the higher magnetic field categories are found at the edges of the structures, while a line of pixels in the lowest categories runs through their centres,  corresponding to the location where the polarity reverses at the limb. 
In contrast, the centre-view offers a natural distribution, where the centre of the region falls into the strongest field categories, and the edges into the weaker ones. These differences in categorisation are consistent with Fig.~\ref{Fig:PixelStatistics}, which shows that a single viewpoint leads to an over-population of the high magnetic field categories at the limb.

The differences shown in Fig.~\ref{Fig:Recategorisation} result from the expansion of magnetic structures with height, which causes field lines to deviate from the radial direction at the edges of facular regions. Although these fields appear weaker at disc centre, their inclination enhances the Stokes $V$ signal near the limb, resulting in higher $B_{LOS}$ values than a purely radial field would. Dividing by $\mu$ then amplifies these values further, leading to the misinterpretation of such pixels as high magnetic field regions at the limb.

For more insight into the differences between the single- and double-
viewpoints, an intermediate step between the two curves of Fig.~\ref{Fig:CurvesVsMu} is presented and discussed in Appendix~\ref{App:A0}.

\section{Conclusions}
The SO/PHI instrument,  when combined with assets on Earth, or orbiting Earth, provides unprecedented opportunities to study photospheric magnetic structures from two viewpoints. 
A first demonstration of the advantages of a second viewpoint in the study of faculae was presented by \citealt{Albert2023Faculae}, who combined data from the SO/PHI-FDT and SDO/HMI at a $60^\circ$ line-of-sight separation. That study showed not only that a second viewpoint enables the observation of facular contrast closer to the limb, but also that systematic differences arise between single- and double-viewpoint results near the limb, mainly because magnetic field measurements are less reliable close to the limb.

We extend the analysis of \citealt{Albert2023Faculae} to high-resolution SO/PHI-HRT observations taken as Solar Orbiter emerged from behind the Sun from Earth's viewpoint, approaching the Sun-Earth line, approximately co-rotating with the Sun. SO/PHI-HRT followed a facular region observed roughly at disc centre from the SO/PHI perspective. The same region, as the Sun rotated into view, appeared at different angles in SDO/HMI, ranging from the limb to $\mu \sim 0.8$. Our study has two main components: (1) a direct comparison of the facular observations near the limb (as seen by SDO/HMI) and at disc centre (as observed by SO/PHI-HRT), and (2) an analysis of the facular contrast as a function of the associated $B_{\mathrm{LOS}}$ and $\mu$ obtained by combining the two viewpoints. The results thus obtained are compared with those from just a single viewpoint.

Comparing the $B_{LOS}/\mu$ of the region measured at the limb by SDO/HMI with that measured at disc centre by SO/PHI-HRT and re-projected to the limb, we find that facular regions at the limb exhibit a polarity reversal in their line-of-sight magnetic fields, consistent with the findings of \citealt{Pietarila2010Expansion}. Here we have verified this behaviour for the first time from a second viewpoint. This is best explained by the expansion of the magnetic field of individual unresolved faculae with height. Groups of such faculae form a “facular bouquet” configuration (\citealt{Pietarila2010Expansion}). In this configuration, flux tubes at the periphery of facular areas (adjacent to quiet Sun regions) are more inclined than those at the centre of the facular complex. This large-scale organisation is discernible in the observations and gives rise to the observed polarity reversal, when viewed at large angles.
As a result of the expansion with height, most of the magnetic field lines in facular regions are not directed normally to the local solar surface, even if on average that is the case. Consequently, the line-of-sight component of the magnetic field carries insufficient information to describe the magnetic field in faculae \citep[see also,][for a more detailed discussion]{Leka2017BLOStoBr}. In principle, a similar analysis could be performed using the strength of the vector magnetic field ($|\vec B|$) to mitigate biases at the limb; however, the higher noise levels in $|\vec B|$ considerably limit the applicability of this approach.

As a consequence of the differences in the derived magnetic field, facular identification methods based on $B_{LOS}$ produce substantially different facular maps from disc-centre and limb perspectives. In the first observation of our time series (corresponding to the largest viewpoint separation) we find that fewer than half ($44\%$) of the faculae identified at disc centre are successfully detected at the limb. Many faculae associated with weaker magnetic fields are missed ($56\%$ of disc-centre facular pixels), while false positives also occur, accounting for $27\%$ of the disc-centre facular map when re-projected to the limb. These false positives falsely associate a significant amount of non-facular signal with magnetic features in single-viewpoint near-limb intensity contrast studies.

Comparing the facular intensity contrast curves derived from the double-viewpoint approach (using SO/PHI-HRT and SDO/HMI) to those obtained from a single viewpoint (using SDO/HMI only), we find that the facular intensity contrast towards the limb is systematically higher in the double-viewpoint case, particularly for faculae associated with weaker magnetic fields. Additionally, we observe that the angle at which the maximum contrast ($\mu_{max}$) occurs in the double-viewpoint approach is located closer to the limb. However, the statistics of the current study are too limited to reliably determine the value of $\mu_{max}$. 
These findings differ from the results of \citealt{Albert2023Faculae}, likely due to several factors: (1) the lower resolution of the SO/PHI-FDT data set used in that study; (2) the fact that the reference "disc-centre" observations in that study spanned a wide $\mu$ range ($0.4 < \mu < 1$), thereby still incorporating limb effects; (3) the derivation of $\mu_{\mathrm{max}}$ being based on a combination of single- and double-viewpoint observations (above and below $\mu = 0.4$, respectively). The main limitation of the present study, however, remains the restricted statistics. Since we observed only a single region over several days, the dataset lacks diversity of structures.

\begin{acknowledgements}
We thank Aimee Norton and Jeneen Sommers for fast-tracking the processing of the PSF-deconvolved data products from SDO/HMI for our study. K.A. thanks Juan Sebastián Castellanos Durán for the many stimulating discussions during this study. Solar Orbiter is a space mission of international collaboration between ESA and NASA, operated by ESA. We are grateful to the ESA SOC and MOC teams for their support. This project has received funding from the European Research Council (ERC) under the European Union's Horizon 2020 research and innovation programme (grant agreement No. 101097844 — project WINSUN). The German contribution to SO/PHI is funded by the BMWi through DLR and by MPG central funds. The Spanish contribution is funded by AEI/MCIN/10.13039/501100011033/ (RTI2018-096886-C5, PID2021-125325OB-C5, PCI2022-135009-2) and ERDF “A way of making Europe”; “Center of Excellence Severo Ochoa” awards to IAA-CSIC (SEV-2017-0709, CEX2021-001131-S); and a Ramón y Cajal fellowship awarded to DOS. The French contribution is funded by CNES. The SDO/HMI data are courtesy of NASA/SDO and the HMI Science Team.
\end{acknowledgements}

\bibliographystyle{aa}
\bibliography{bibfile.bib}

\begin{appendix}

\section{Re-projection of SO/PHI facular map onto SDO/HMI's coordinate system}\label{App:B}
\begin{figure}[h!]
   \centering
    
    \includegraphics[width=\hsize]{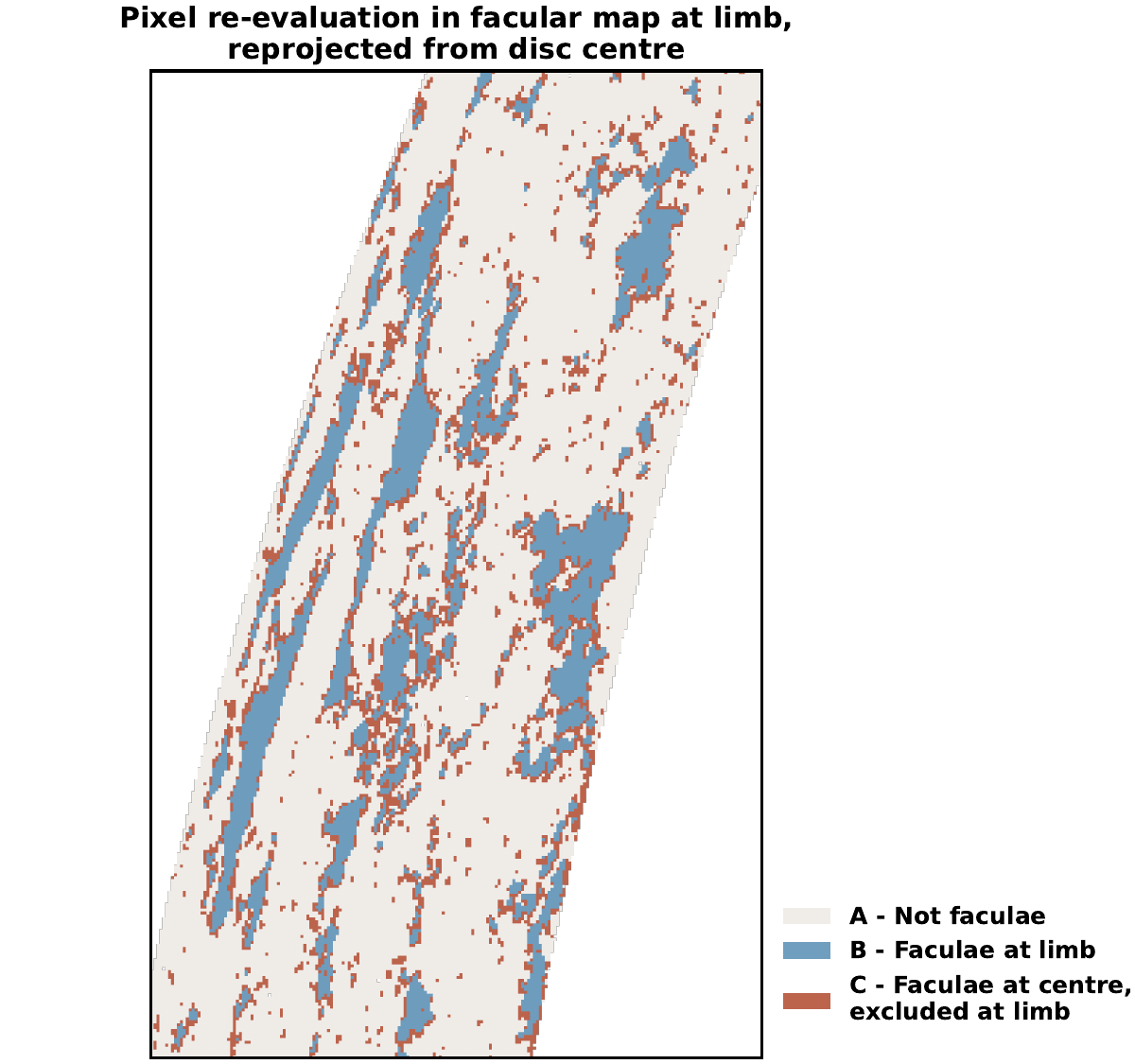}
    
     \caption{A sub-region of the first dataset in this study (same as in Fig. \ref{Fig:BLOS_compared}), showing the facular map determined at disc centre by SO/PHI-HRT and reprojected onto the reference frame of SDO/HMI. After re-projection, an additional selection criterion is applied: only those facular pixels are kept, for which the reprojected $B_{LOS}/\mu$, originally measured at disc centre, exceeds three times the noise level of SDO/HMI at the limb. Pixels meeting this criterion are shown in colour B, while those excluded from the analysis are shown in colour C.
     }
         \label{Fig:Faculae_reproj}
   \end{figure}
When reprojecting facular maps from disc centre to the limb, the original map compresses into fewer pixels. As a result, each pixel in the reprojected map contains varying degrees of facular contribution. Pixels with minimal facular contribution may not exhibit facular-like intensity behaviour at the limb.

To separate pixels with sufficient facular contribution from those without, we apply a threshold criterion. Specifically, pixels must also be classified as faculae at the limb based on their reprojected $B_{LOS}$. To ensure that this is the case, as a conservative approach, we require the reprojected $B_{LOS}$ of any facular pixel to exceed three times the noise level (denoted $\sigma$) of the limb observation (consistently with our single-viewpoint analysis): 
\begin{equation}
    B_{LOS,reproj.}>3\sigma_{B_{LOS,limb}};
\end{equation}
otherwise we exclude these pixels from the analysis.

Figure \ref{Fig:Faculae_reproj} shows the reprojected facular map, where pixels accepted as faculae are shown in blue (colour B), while those excluded from the analysis are in orange (colour C). The exclusion of these pixels impacts only panel (h) in Fig. \ref{Fig:CurvesVsMu} -- including them would lower the contrast close to the limb for this panel.

\section{Approaching the limb}\label{App:A}
\begin{figure}[h!]
   \centering
    
    \includegraphics[width=.75\hsize]{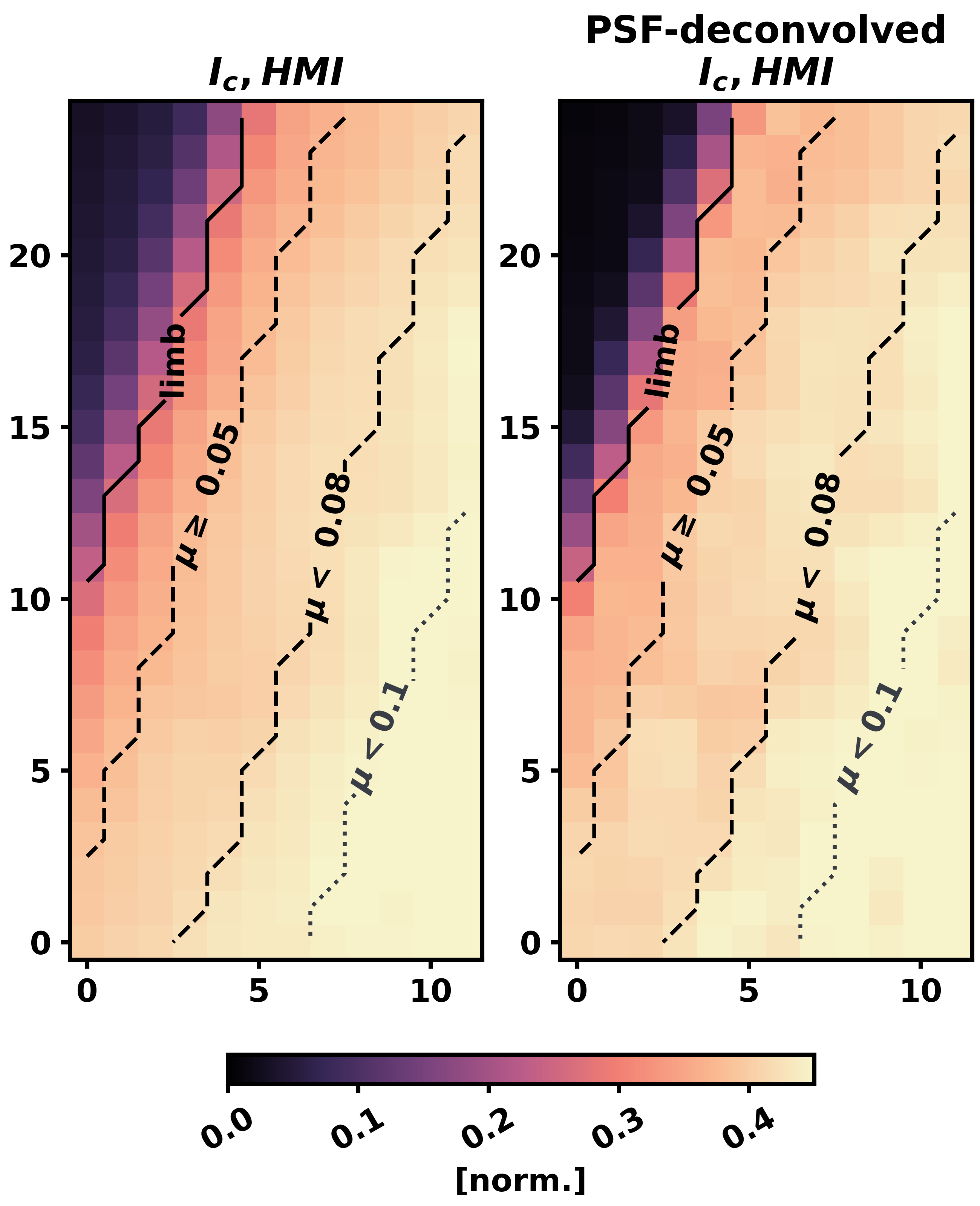}
    
     \cprotect \caption{Close-up of SDO/HMI continuum intensity pixels near the solar limb. Left: Continuum intensity from the \verb|hmi.Ic_45s| data product, normalised to the mean intensity at disc centre. Right: Corresponding normalised intensity after PSF deconvolution, from the \verb|hmi.Ic_45s_dcon| data product. The associated $\mu$ values are computed from the metadata of the \verb|hmi.Ic_45s| data product. The solid line marks the solar limb; dashed lines indicate the outermost bin-width (used in Fig.~\ref{Fig:CurvesVsMu}); and the dotted line shows the $\mu=0.1$ contour for reference.
     }
         \label{Fig:BinWidths}
  \end{figure}

\begin{figure*}[h!]
   \centering
   
    \includegraphics[width=\hsize]{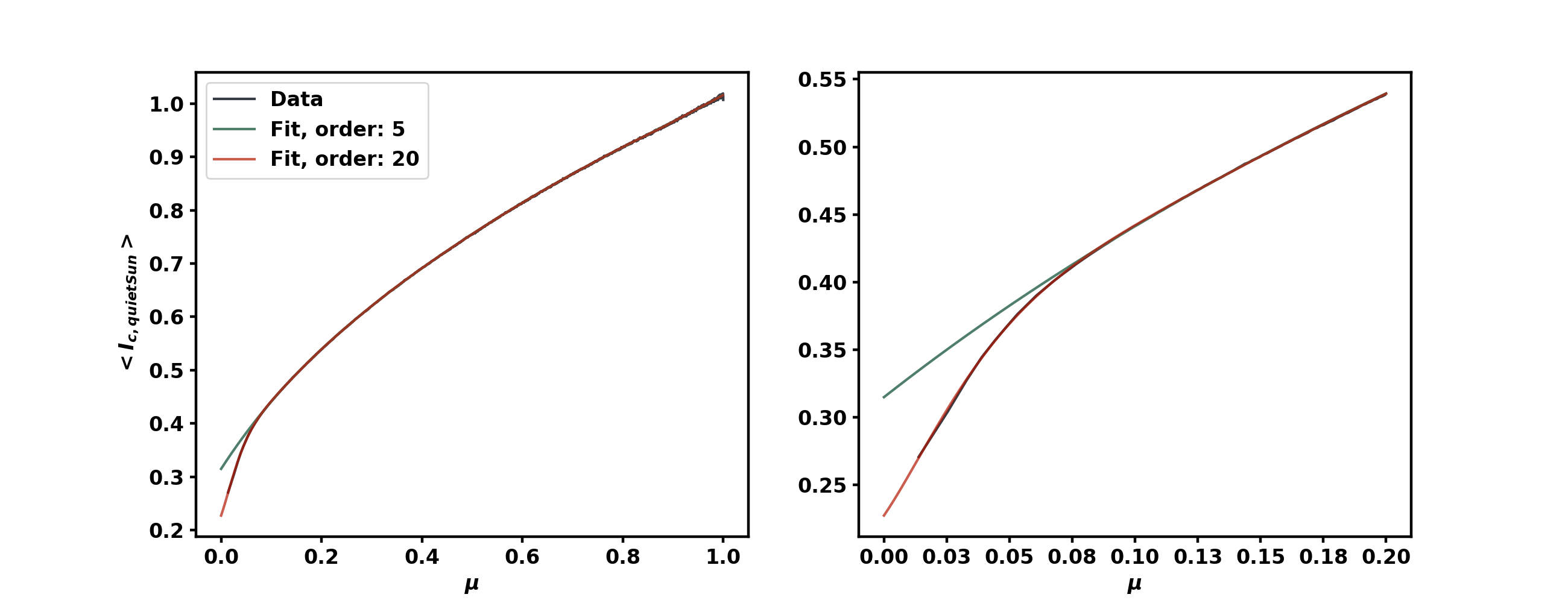}

     \cprotect \caption{The centre-to-limb variation (CLV) of the quiet Sun intensity in the first \verb|hmi.Ic_45s| data set used in the study, along with fifth- and twentieth-order polynomial fits. The left panel shows the full solar disc, while the right panel zooms in on the range $0 < \mu < 0.2$.
     }
         \label{Fig:CLV}
   \end{figure*}
Profiting from the double view-point, we aim to analyse data closer to the limb than previous studies did -- which typically consider pixels with $\mu>0.1$. There are two important considerations in this regard: (1) how close are these pixels to the solar limb; and (2) how well can we reconstruct the centre-to-limb variation of the intensity in these pixels.

In this study, we analyse data as close to the limb as $\mu = 0.05$. The first bin in Fig. \ref{Fig:CurvesVsMu} includes pixels with $0.05\le\mu<0.08$. Figure \ref{Fig:BinWidths} shows where these pixels fall in relation to the solar limb, and in relation to the more traditional $\mu>0.1$ approach. The colour scheme shows the $I_c$ at the SDO/HMI limb, for both the \verb|hmi.Ic_45s| data product and its PSF deconvolved counterpart, the \verb|hmi.Ic_45s_dcon|.
The points considered in our analysis are all at least two pixels away from the solar limb -- we consider the solar limb to be where it is defined in the metadata of the SDO/HMI data products. This region also avoids the pixels where the PSF clearly has a very strong influence (which is the case for the closest one to two pixels to the limb).

The use of the \verb|hmi.Ic_45s| data product has the drawback that centre-to-limb variation (CLV) corrected continuum intensity data are not provided. Previous studies \citep[e.g.,][]{Albert2023Faculae, Yeo2013Intensity} apply a fifth-order polynomial fit to the CLV of the quiet sun intensity, following the method of \citet{neckel_solar_1994}, considering only points with $\mu > 0.1$ in their analyses. 

However, in the present study we derive facular intensity curves as close to the limb as $\mu < 0.05$. Since the intensity contrast of magnetic elements is defined in relation to the local $\mu$-dependent mean quiet Sun intensity, the accurate determination of this reference is fundamental to our analysis. We have, therefore, evaluated the accuracy to which this relationship is determined, with special focus on the regions near the extreme limb.

Figure~\ref{Fig:CLV} shows the mean quiet Sun intensity as a function of $\mu$ for the first SDO/HMI data set used in the study (gray). The mean quiet Sun intensity is derived from pixels with $B_{LOS}/ < 2*\sigma$, as a conservative threshold, where $\sigma$ is the noise level of the SDO/HMI $B_{LOS}$, determined as a function of $\mu$ (see also Sec. \ref{Sec:facularMap}). These pixels are sorted by their associated $\mu$ values and grouped into bins of $5000$ pixels each, for which the mean is calculated. We then fit the resulting profile with both a fifth-order and a twentieth-order polynomial (green and red, respectively).

The fifth-order fit overestimates the intensity for $\mu < 0.08$, causing an artificial decrease in the derived intensity contrast in this region. Although the sudden intensity drop in the CLV profile is likely instrumental, using a higher-order polynomial substantially improves the fit. For this study, we adopt the twentieth-order polynomial, which provides a very good match down to $\mu \approx 0.035$, as shown in Fig. \ref{Fig:CLV}. This allows us to confidently include points with $\mu > 0.05$ in our analysis.

\section{Differences between the results from a single and double view-points}\label{App:A0}

\begin{figure*}[h!]
   \centering
    \includegraphics[width=\hsize]{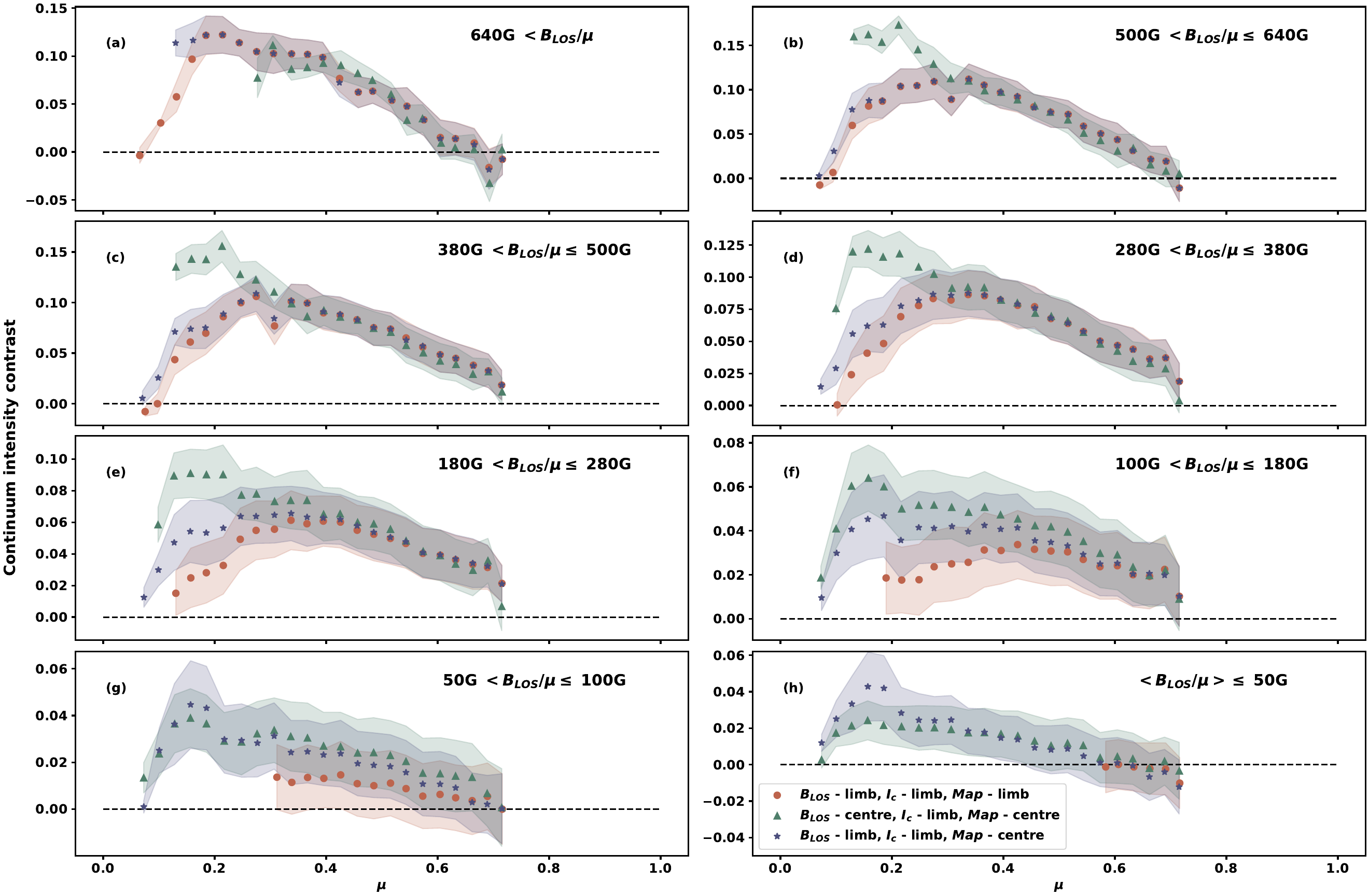}
    
     \caption{Same relationship as shown in Fig.~\ref{Fig:CurvesVsMu}, but limited to data in the common field-of-view of SDO/HMI and SO/PHI, which falls between $0.05 <\mu < 0.75$. Consequently, the orange curve now uses much less data than that shown in Fig.~\ref{Fig:CurvesVsMu}, and therefore has more limited statistics, but it is calculated in the same way as the one in Fig.~\ref{Fig:CurvesVsMu}. The green curve (double-viewpoint approach) remains unchanged. An additional blue curve is included, representing the facular contrast derived from both $I_c$ and $B_{\mathrm{LOS}}$ measured at the limb, but using pixels selected from the facular map derived at disc centre. Markers and shaded areas have the same meaning as in Fig.~\ref{Fig:CurvesVsMu}.
     }
         \label{Fig:SameRegion}
  \end{figure*}
  
For additional insight, we present an intermediate step between the green and orange curves in Fig.~\ref{Fig:CurvesVsMu}. Figure \ref{Fig:SameRegion} shows three curves: the green curve is the same as the one in Fig.~\ref{Fig:CurvesVsMu}, while the orange curve is from SDO/HMI-data only, however it is restricted to the region which is seen by both SDO/HMI and SO/PHI, instead of using data from the full solar disc (as in Fig.~\ref{Fig:CurvesVsMu}). The intermediate curve, in blue, is calculated based on pixels identified as faculae at the disc centre, while retaining the $B_{\mathrm{LOS}}/\mu$ values measured at the limb (consistent with the single-viewpoint approach). See also the legend in Fig.~\ref{Fig:SameRegion} for an overview of the differences between the curves. 
  
We can now examine the differences and agreements between these curves. First, we focus on the comparison of the orange and blue curves, which represent, respectively, the purely single-viewpoint data and the data where the facular map is derived from the disc-centre perspective. As expected from the increase of noise and decrease of resolution at the limb, most of the additional pixels identified by the new facular map correspond to low magnetic field strengths (see also Sec. \ref{Sec:facularMap}). This improved identification of faculae near disc centre effectively extends the curve closer to the limb for the lowest magnetic field categories (panels f to h). 

Faculae with stronger magnetic field are identified in both viewpoints, and as a result the blue and orange curves are close in panels a to d. However, in panel a, while the orange curve extends all the way to $\mu = 0.05$, the blue curve stops at $\mu = 0.11$. This suggests that at the extreme limb, the single-viewpoint identification introduces several high-magnetic-field pixels that do not correspond to faculae. These may result from measurement noise, amplified by the division by $\mu$, or from strong Stokes $V$ signals produced by horizontal magnetic fields in the quiet Sun.

Second, we examine how the results change when incorporating disc-centre information about the magnetic field. We compare the blue and green curves: the blue curve uses the facular map from disc centre but retains $B_{\mathrm{LOS}}$ values from the limb, while the green curve uses both the facular map and magnetic field values from disc-centre observations. In this comparison, a key consideration is the reclassification of pixels into different magnetic field categories, as changes in $B_{\mathrm{LOS}}/\mu$ can shift pixels between panels. These changes can be interpreted in terms of change in the contrast curves, under the general assumption that faculae associated with stronger magnetic fields tend to exhibit higher contrasts. 

For example, in panel (h) of Fig.~\ref{Fig:SameRegion}, the polarity reversal line (see Fig.~\ref{Fig:Recategorisation}) appears as a peak in the contrast in the blue curve just below $\mu=0.2$. It is the results of high-intensity pixels near the solar limb being incorrectly assigned very low magnetic field values. When these pixels are reassigned to the correct magnetic field categories using disc-centre information, this contrast enhancement disappears in the green curve. In panel (g), which also represents low magnetic fields, the blue and green curves remain in good agreement. This suggests that although some pixel exchange between magnetic field categories occurs, the majority of the contributing pixels are newly introduced and are associated with low magnetic fields at the limb as well.

At higher magnetic field strengths (panels a to f), the contrast in each category increases when using disc-centre information. 
This trend reflects the general overestimation of $B_{\mathrm{LOS}}/\mu$ near the limb, as discussed in Sec.\ref{Sec:resultsViewPoints}, and illustrated in Fig.\ref{Fig:Recategorisation}. After correction, pixels that were previously misclassified into higher field categories move into lower ones; and in panel a, no pixels remain below $\mu = 0.23$. This re-sorting also increases the mean intensity contrast: in the high-field categories by removing lower-contrast, misclassified pixels, and in the lower-field panels by correctly adding higher-contrast pixels.
\end{appendix}

\end{document}